\documentclass[%
10pt,%
onecolumn,%
oneside,%
floats,%
aps,%
prd,%
nobibnotes,%
nofootinbib,%
amsmath,%
amssymb,%
amsfonts,%
superscriptaddress,%
eqsecnum,%
notitlepage,%
]{revtex4-1}

\usepackage[utf8]{inputenc}
\usepackage{graphicx,array,dcolumn,cases}

\usepackage{hyperref}
\hypersetup{hidelinks}

\def\mb{\mathbf}
\def\be{\begin{eqnarray}}
\def\ee{\end{eqnarray}}
\def\mpl{m_{Pl}}

\def\mn{{\mu\nu}}

\def\-g{\sqrt{-g}}

\newcommand{\rar}{\rightarrow}

\renewcommand\rho{\varrho}
\renewcommand\tilde{\widetilde}

\def\so{\Rightarrow}

\newcommand{\StartAgainHere}{\vspace*{2cm}
\begin{center}
 * * * * * * * * * * * 
\end{center}
\vspace*{2cm}
}

\begin{document}

\title{Gravitational instability in oscillating background}

\author{E.V. Arbuzova}
\email{arbuzova@uni-dubna.ru}
\affiliation{Novosibirsk State University, Novosibirsk, 630090, Russia}
\affiliation{Department of Higher Mathematics, University "Dubna", 141980 Dubna, Russia}

\author{A.D. Dolgov}
\email{dolgov@fe.infn.it}
\affiliation{Novosibirsk State University, Novosibirsk, 630090, Russia}
\affiliation{ITEP, Bol. Cheremushkinsaya ul., 25, 113259 Moscow, Russia}
\affiliation{Dipartimento di Fisica e Scienze della Terra, Universit\`a degli Studi di Ferrara\\
Polo Scientifico e Tecnologico - Edificio C, Via Saragat 1, 44122 Ferrara, Italy}

\author{L. Reverberi}
\email{reverberi@fe.infn.it}
\affiliation{Dipartimento di Fisica e Scienze della Terra, Universit\`a degli Studi di Ferrara\\
Polo Scientifico e Tecnologico - Edificio C, Via Saragat 1, 44122 Ferrara, Italy}
\affiliation{Istituto Nazionale di Fisica Nucleare (INFN), Sezione di Ferrara\\
Polo Scientifico e Tecnologico - Edificio C, Via Saragat 1, 44122 Ferrara, Italy}

\begin{abstract}
Evolution of density and metric perturbations in the background of high frequency oscillations of curvature in $F(R)$ gravity is considered. In addition to the usual Jeans-like instability new effects of amplification of perturbations, associated with parametric resonance and antifriction phenomena, are found. 
\end{abstract}

\maketitle

\section{Introduction \label{s-intro}}

Gravity modifications at large distances have been suggested~\cite{grav-mod-1} for an explanation of the accelerated cosmological
expansion observed in the present-day universe. The idea is to add a non-linear function of curvature scalar, $F(R)$,
to the standard Einstein-Hilbert action: 
\be
A_\text{grav} =  -\frac{m_{Pl}^2 }{16 \pi} \int d^4 x \sqrt{-g} \,\left[ R + F(R) \right]\,,
\label{A-grav}
\ee
such that the modified Einstein equations have an accelerated De Sitter-like solution in the absence of matter\footnote{We use natural units $c = \hbar = 1$, and Newton's constant is defined as $G_N = \mpl^{-2}$. The metric signature is $(+,-,-,-)$, the Riemann tensor is defined as $R^\alpha_{\,\mu\beta\nu} = \partial_\beta\gamma^\alpha_\mn + \cdots $, and the Ricci tensor is $R_\mn = R^\alpha_{\,\mu\alpha\nu}$.}. These theories contain an additional massive scalar degree of freedom (dubbed ``scalaron'') beside the usual massless, spin-2 graviton. In the first works~\cite{grav-mod-1} the function $F(R)$ was taken in the form $F(R) = - \mu^4 /R$, where $\mu$ is a constant parameter with
dimensions of a mass and of the same order of magnitude as the present Hubble parameter (or inverse universe age), i.e. $\mu^2 \sim | R_c | \sim 1/t_U^2$, where $t_U \approx 14 $ Gyr is the universe age. It was found~\cite{ad-mk},
however, that in astronomical systems with even slightly varying mass density a very strong instability would develop, resulting in explosive solutions evidently incompatible with observations.

To cure this shortcoming further modifications of GR have been suggested~\cite{Starob,HuSaw,ApplBatt}, for a review see Ref.~\cite{mod-grav-rev}. Let us take as a guiding example the specific $F(R)$ of Ref.~\cite{Starob}:
\be
F(R) =  -\lambda R_c\left[1-\left(1+\frac{R^2}{R_c^2}\right)^{-n}\right] , 
\label{F-AS}
\ee
where $R_c$ is a constant parameter with dimension of curvature and close by magnitude to the present cosmological curvature, $\lambda$ is a dimensionless constant of order unity and the power $n$ is usually taken to be an integer (though not necessarily so). This choice of $F(R)$ leads to a further problem, namely the solution of the equation of motion for the gravitational field must be singular with $R\to \infty$ in the past to produce a reasonable late-time cosmology~\cite{App-Bat-Star}. Similarly, it has been found that systems with rising mass/energy density will evolve to a $R\to\infty$ singularity 
in the future~\cite{future-sing}. The singularity can be avoided if one adds an extra term $F(R) \rar F(R) - R^2/(6m^2)$~\cite{App-Bat-Star}. At small $R$ the system tends to evolve to higher values of curvature, but as $|R|$ grows the $R^2$-term eventually becomes dominant and pushes the system back to lower curvatures. This results in oscillating solutions $R(t)$, possibly with very large amplitude. In our papers \cite{ADR-1} we have found such oscillating solutions for the particular example of $F(R)$ given by Eq.~(\ref{F-AS}) with the addition of the $R^2$ term. The properties of different versions of $F(R)$ are such that they induce effective potentials for $R$ having minimum at $R$ equal to its GR value, $R=R_\text{GR}$, and rising in both directions of lower and higher $R$. So an oscillating behaviour of $R$ is a generic feature of these modified gravity theories. We assume that the form of such oscillations is arbitrary, keeping their amplitude and frequency as free parameters. 

Gravitational instability in modified gravity has been considered in Refs.~\cite{instab-mod}. In our paper \cite{ADR-Jeans} we studied 
the Jeans instability in classical and modified gravity in a background with rising energy density, assuming that the
background metric changes slowly as a function of space and time compared to the typical frequency and Compton wavelength of the scalaron field.
Here we investigate gravitational instability in a quickly oscillating curvature background.
The time evolution of first-order perturbations is governed by a fourth-order 
differential equation (instead of the usual second order one), 
therefore new types of unstable solutions are expected, and have in fact been found.

There appear not only the usual Jeans solution with a slightly modified (reduced) length scale, but also parametric resonance amplification of density perturbations, and an amplification of the perturbations due to the
 "antifriction" behaviour of the coefficients of odd derivatives (first and third) in the equation.

The paper is organized as follows. In Section~\ref{s-F-of-R} we give a general description of density perturbations in a quickly 
oscillating background; in Section~\ref{sec-Jeans} we discuss the Jeans-like instability in modified gravity. In 
Section~\ref{sec-param_res} parametric resonance amplification of fluctuations is studied for harmonic and spiky oscillations 
of the background curvature. In Section~\ref{s-antifric} we investigate the effects of the antifriction induced by the change in the signs of the coefficients in front of the odd derivatives in the equation of motion. Finally, Section~\ref{s-conclusions} is dedicated to the discussion of the results and  possible implications. In Appendix~\ref{s-metric} we present the expressions for the metric, Christoffel symbols, Ricci and Einstein tensors for a Schwarzschild-like isotropic and homogeneous solution in modified gravity.

\section{Density perturbations over a quickly varying background }
\label{s-F-of-R}

\subsection{Basic framework and equations }

The modified Einstein field equations, derived from the action (\ref{A-grav}), are:
\be
\left( 1 + F'_R\right) R_{\mu\nu} -\frac{1}{2}\left( R + F\right)g_{\mu\nu}
+ \left( g_{\mu\nu} D_\alpha D^\alpha - D_\mu D_\nu \right) F'_R  = 
\frac{8\pi T^{(m)}_{\mu\nu}}{m_{Pl}^2} \equiv  \tilde T_{\mu \nu} ,
\label{eq-of-mot}
\ee 
where $F'_R = dF/dR$, $D_\mu$ is the covariant derivative and $T^{(m)}_{\mu\nu}$ is the energy-momentum tensor of matter.

We assume that the background metric weakly deviates from the Minkowsky one, while derivatives of the metric may be far from their GR values. In particular, $R$ may be very much different from $R_\text{GR} = -\tilde T$. Please note that these two conditions are only apparently contradictory; in fact $R \sim \partial^2 g$ so even if the metric is approximately the Minkowsky one, the curvature may be very large in comparison with $R_{GR}$, 
e.g if the frequency of the oscillations of $g$ and $R$ is high, as in the case under scrutiny.
We consider astronomical systems with $|R_c| \ll |R| \ll m^2$. Both limits are natural for relatively dense systems, that is denser than the average cosmological background but much less dense than $m_{Pl}^2 m^2$. Note that $m \gtrsim 10^{5}$ GeV, see e.g.~\cite{ADR-cosm}. In this limit we have
\be\label{approx_conditions}
|F(R)| \ll |R|\,,\, |F'(R)| \ll 1\,.
\ee
This is surely fulfilled for the model~(\ref{F-AS}), for which at $R\gg R_c$:
\be
F(R) \approx -\lambda R_c \left[ 1 -\left(\frac{R_c}{R}\right)^{2n} \right] - \frac{R^2}{6 m^2} \,.
\label{F-large-R}
\ee
Therefore, equation (\ref{eq-of-mot}) can be approximated as:
\be
G_{\mu\nu} + \frac{1}{3\omega^2}(D_{\mu}D_{\nu} - g_{\mu\nu}D^2)R = \tilde T_{\mu \nu}\, ,
\label{EoM-F(R)}
\ee 
where $G_{\mu\nu} = R_{\mu\nu} - g_{\mu\nu} R/2$ is the Einstein tensor and 
\be
\omega ^{-2} = - 3 F''_{RR}\,.
\label{omega-F} 
\ee
Once written in this form, the equation is largely independent of the specific model considered except of course for the value of $\omega$, and provided that along the background solution $\omega \simeq $ const., or rather $\dot\omega/\omega^2 \ll 1$, and that the conditions~(\ref{approx_conditions}) are fulfilled. In the example
(\ref{F-large-R}):
\be
\omega^2 = \left[\frac{1}{m^2} + \frac{6\lambda n (2n+1)}{|R_c|}\, \left( \frac{R_c}{R}\right)^{2n +2} \right]^{-1}\,.
\label{omega-2}
\ee
Cosmological perturbations started to rise at the onset of the matter dominated epoch at redshift $z_{eq} \simeq 10^4$ when 
$R_{eq}/R_c \sim 10^{12}$. So for $m=10^5$ GeV, which is its lower limit~\cite{ADR-cosm}, $\omega$ may be treated as practically 
constant
if $n\gtrsim 3$. For systems with energy density of about 1 $\rm{g/cm^3}$ the frequency $\omega$ would remain constant even for $n\geq 1$.
If $\omega $ rises with time, the perturbations would rise even faster than obtained below. If~(\ref{omega-2}) is dominated 
by the second term in square brackets, that is
\be
\omega^2 \approx \frac{|R_c|}{6\lambda n (2n+1)}\, \left( \frac{R}{R_c}\right)^{2n+2}\,,
\label{om-2}
\ee
then the frequency might depend crucially on time and the approximation of constant frequency would be invalid, although $|R| \gg |R_\text{GR}|$ makes the second term in Eq.~(\ref{omega-2}) smaller. This corresponds to the results shown in Ref.~\cite{ADR-1}, in which we found quickly oscillating solutions with $R$ strongly deviating from $R_\text{GR}$.

We consider a spherically symmetric cloud of matter with initially homogeneous energy density inside the limit 
radius $r_m$. We choose the Schwarzschild-like isotropic coordinates in which the metric takes the form
\be
ds^2 = A\,dt^2 + B\,\delta_{ij}\,dx^i dx^j\,,
\ee
where $A$ and $B$ are functions of space and time. The corresponding expressions for the Christoffel symbols and Ricci tensor can be found in Appendix~\ref{s-metric}. As usual the metric and the curvature tensors are expanded around their background values at first order in infinitesimal perturbations, i.e.
\be
\begin{aligned}
A & = A_b + \delta A\,,\\
B & = B_b + \delta B\,,\\
R & = R_b + \delta R\,.
\end{aligned}
\ee
Since $A_b$ and $R_b$ are quickly oscillating functions of time with possibly large amplitude ("spikes") which were found in our previous works and their time derivatives could be large, we shall keep terms of the second order in $\partial_t$, such as $\partial^2_t A$, $\partial_t A\,\partial_t R$, and so on. 

As it was done previously in our paper \cite{ADR-Jeans}, we describe the evolution of perturbations using the equation for $G_{tt}$, the $\partial _i \partial _j$-component of the equation for $G_{ij}$, and the Euler and continuity equations which follow from the covariant conservation conditions $D_{\mu} T^{\mu}_j=0$ and $D_{\mu} T^{\mu}_t=0$. However, in the present work we do not confine ourselves to a static background metric, but take into account its time variation which leads to several new effects.

\subsection{First order perturbations}
\label{sec-first_order_perturb}

The $tt$-component of Eq.~(\ref{EoM-F(R)}) is:
\be 
-\frac{\Delta B}{B^2} + \frac{1}{3\omega^2B} \left (\Delta R - \frac{3\dot B \dot R}{2A} + \frac{\partial^iB\,\partial_i R}{2B}\right ) = \tilde \rho\,,
\label{tt-mod}
\ee
because, according to (\ref{G-tt}) and (\ref{T-tt}), $G_{tt}= - A\Delta B/B^2$ and $\tilde T_{tt} = \tilde \rho A$. We take the background as weakly dependent on space coordinates but quickly oscillating with time, so that we can neglect the last term in the l.h.s. of Eq.~(\ref{tt-mod}). The corresponding equation for perturbations takes the form:
\be
-\Delta \delta B - 2\tilde\rho_b\,\delta B + \frac{1}{3\omega^2} \left (\Delta \delta R - \frac{3}{2} \delta \dot B\dot R_b\right ) = \delta \tilde \rho\,,
\label{tt-fluct}
\ee
where $\delta R$ is [see (\ref{R})]: 
\be
\delta R= \Delta \delta A - 3 \delta \ddot B + 2\Delta \delta B + \frac{3}{2} \dot A_b\,\delta \dot B\, . 
\label{R-fluc}
\ee
We assume that $\dot B_b$ is small in comparison with $\dot A_b$ and $\dot R_b$ and put $A_b=B_b=1$ in denominators, see Eqs.~(\ref{B-of-r-t}) and (\ref{A-of-r-t}). Analogously we find the $\partial _i \partial _j$-component of the equation for $G_{ij}$:
\be
\partial _i \partial _j\left(\delta A+\delta B - \frac{2\delta R}{3\omega^2}\right)=0\,,
\label{ij-fluct}
\ee
which remains unchanged with respect to Ref.~\cite{ADR-Jeans}. The continuity equation, derived from $D_{\mu} T^{\mu}_t=0$, is not changed either:
\be 
\delta\dot \rho  + \rho_b\, \partial_j  U^j + \frac{3}{2}\rho_b\,\delta \dot B=0\,,
\label{cont-fluc}
\ee
while the Euler equation acquires an additional term $\dot A_b \, \rho _b\, U_j/2$:
\be 
\rho_b\, \delta \dot U_j + \partial_j  P + \frac{1}{2}\rho_b \,\partial_j \delta A  + \frac{1}{2} \dot A_b \, \rho _b\, U_j=0\,.
\label{Euler-fluc} 
\ee
Such term obviously vanishes in static background, therefore it had been previously neglected. Notice that despite quick oscillations of $A_b$ and $R_b$, the background energy density varies slowly with time, in agreement with the 
solution found in our paper~\cite{ADR-1}.

Introducing $U^j= - U_j = - \partial_j \sigma$, $P=c_s^2 \delta \rho $ and looking for the solution in the form $\sim \exp [ - i\bf k \cdot \bf x]$, we obtain the following system of equations for the five unknown functions of time, $\delta A$, $\delta B$, $\delta R$, $\delta \tilde \rho$, and $\sigma $:
\begin{subnumcases}{\label{system_full}}
\delta\tilde\rho = (k^2 - 2\tilde\rho_b)\delta B - \frac{k^2}{3\omega^2}\,\delta R - \frac{\dot R_b}{2\omega^2}\,\delta\dot B\,,\\
\delta R = \frac{3}{2} \omega ^2 (\delta A + \delta B)\,, \\
\delta R = - k^2( \delta A + 2 \delta B )- 3 \delta \ddot B  + \frac{3}{2} \dot A_b \delta \dot B\,, \\
\delta \dot {\tilde \rho} + \tilde \rho_b\, k^2 \sigma + \frac{3}{2}\tilde \rho_b\,\delta \dot B=0\,, \\
\tilde \rho_b \dot \sigma - c_s^2 \delta \tilde \rho + \frac{1}{2} \tilde \rho_b (\dot A_b \sigma - \delta A) = 0\, . 
\end{subnumcases}

At this stage we should comment on the application of the Fourier transformation to the derivation of equations~(\ref{system_full}). The Fourier transform over space variables can be applied straightforwardly only if the coefficients in these equations are space-independent. It might be so in physically interesting cases, however in our example the metric function $A_b$ depends explicitly upon space coordinates, in fact [see~(\ref{A-of-r-t})]: 
\be\label{eq_Ab_Rb}
A_b = 1 + \frac{R_b r^2}{6}\,.
\ee
Evidently, for sufficiently small $r$, such that $ R_b r^2/6 \ll 1 $, so that we can completely neglect this term, 
  the Fourier transformation  can be safely applied. However, neglecting this term we exclude the effect of time dependent background, which leads to new forms of instability. 

Nevertheless, we can take into account the effects of $r$-dependent terms and justify the applicability of the Fourier transformation both in the limits of large and small $r$. 

In the care of large $r$ the Fourier transformation is applicable in the adiabatic case, when the background quantities change little over one wavelength of perturbations, $\lambda = 2\pi/k$.  Indeed, performing the Fourier transform of the equations essentially consists in multiplying them by 
$\exp (i\mb k \cdot \mb r)$ and integrating over $ d^3 r$. If the coefficients of the linear equations for infinitesimal perturbations do not depend upon space coordinates we come as usual to an algebraic system of linear equations for the Fourier modes of the fluctuations. If, as is our case, some coefficients in the original differential equations depend upon $r$, then we can still transform our equations taking the 
integral $ d^3 r \exp (i\mb k \cdot \mb r)$ not over the whole space but in some neighbourhood of a fixed value $r = r_0$ with 
radius $\Delta r $, chosen so that the space dependent coefficients are practically constant in such neighbourhood. To this end it is 
necessary that the range of integration contain many wave lengths, i.e. $k \Delta r \gg 1$, and at the same time that $\Delta r $ be sufficiently small, 
so that the $r$-dependent coefficients may be considered as approximately constant. Due to the largeness of $k \Delta r$, such an integral would be close to the real Fourier transform with infinite integration limits. 

{This essentially allows us to treat $r$ as a constant parameter. Our final results will of course depend on such coordinate, in fact we will present results obtained in the two different limits $kr \gg 1$ and $kr \ll 1$. The former condition is the more natural one, because the adiabatic approximation essentially implies $k r_m \gg 1$ and most values of $r$ are smaller but roughly of the same order of magnitude of the limit radius $r_m$. The latter condition holds only near the centre of the cloud, that is for relatively small values of $r$; however, this may be a non negligible portion of the total system.} {The case of small $r$ is considered below, after Eq.~(\ref{deltaB-4dots-lin})}.

Let us examine this condition more quantitatively. We demand that perturbations vary quickly compared to the background, that is
\be
\left|\frac{\partial_r A_b}{A_b}\right|,\,\left|\frac{\partial_r^2 A_b}{\partial_r A_b}\right| \ll k\,,
\ee
which, using~(\ref{eq_Ab_Rb}), implies
\be
\label{eq-R_r2_condition}
|R_b| r^2 \ll 1\,,\qquad k r \gg 1\,.
\ee
Hence, choosing wavenumbers $k \gg r_m^{-1}$ (see discussion above) and considering $|R_b| r_m^2 \ll 1$, as is the case for dilute enough systems, 
these conditions are nicely fulfilled and the Fourier expansion is reliable. The mass of the object under scrutiny is 
$M_{tot} = 4\pi \rho_b r_m^3/3$. We express $\tilde \rho_b$ as
\be
\tilde \rho_b \equiv \frac{8\pi \rho_b}{m_{Pl}^2} = \frac{3 r_g}{r_m^3}\,,
\label{rho-of-rg}
\ee
where $r_g = 2M_{tot} /m_{Pl}^2 $ is the gravitational (Schwarzschild) radius of the system. 
Correspondingly~(\ref{eq-R_r2_condition}) becomes
\be
\frac{|R_b|}{\tilde\rho_b} \ll \frac{r_m}{3 r_g}\,,
\ee
which sets an upper value for the amplitude of oscillations at which our approximations fail. Such constraint leaves a large portion of 
solutions as suitable for our analysis; for instance, for a solar mass cloud with mass density $10^{-24}$ g/cm$^3$ the ratio is 
$r_m/r_g \sim 10^{13}$.

\subsection{Evolution equation}

Based on the comments above we derive from the system of five low order equations~(\ref{system_full}) the fourth order equation for the single function $\delta B$:
\be
&&\delta \overset{....}{ B}  -  \left( 1 + \frac{2k^2}{3\omega^2}\right)\frac{\dot R_b}{2k^2} \,\delta \overset{...}{ B} +
\left[ \omega^2 \left(1- \frac{\tilde\rho _b }{2 k^2} \right)  + k^2(1+c_s^2)  - \frac{4 \tilde\rho _b}{3}  
 - \ddot A_b - \frac{1}{k^2}\left(1 + \frac{2k^2}{3\omega^2}\right)\left(\ddot R_b + \frac{\dot A_b\dot R_b}{4}\right) - \frac{\dot A_b^2}{4}  \right] \delta \ddot B +  \nonumber \\
&&+\left[-\frac{\overset{...}{A_b}}{2} - \frac{1}{4k^2}\left(1+ \frac{2k^2}{3\omega^2}\right) \left( 2\overset{...}{R_b} + \dot A_b \ddot R_b + 2\dot R_b c_s^2k^2 \right) - \frac{\ddot A_b\,\dot A_b}{4} + \frac{\dot A_b}{2}
\left( \omega^2 \left(1- \frac{\tilde\rho _b }{2 k^2} \right)  + k^2(1 - c_s^2)  + \frac{2 \tilde\rho _b}{3}   \right)
 \right]\delta \dot B  +  \nonumber \\
&&+\left[ c_s^2 k^2(k^2+\omega^2) - \frac{\tilde \rho_b \omega^2}{2}\left(1+\frac{4k^2}{3\omega^2}\right)  - 2 c_s^2 \tilde \rho_b \omega^2 \left(1+\frac{2k^2}{3\omega^2}\right)
\right] \delta B  = 0\,.
\label{deltaB-4dots-normlz}
\ee 
In a wide parameter range, essentially whenever~(\ref{eq-R_r2_condition}) holds, the following conditions are fulfilled:
\be
\begin{gathered}
  \dot A_b^2 \ll \ddot A_b\,,\\
  \dot A_b\dot R_b \ll \ddot R_b\,,\\
  \dot A_b\ddot A_b \ll \overset{...}{A_b}\,,\\
  \dot A_b\ddot R_b \ll \overset{...}{R_b}\,,
  \label{bilinear}
\end{gathered}
\ee
therefore several terms in Eq.~(\ref{deltaB-4dots-normlz}) can be neglected. Using Eq.~(\ref{eq_Ab_Rb}) we obtain:
\be
&&\delta \overset{....}{ B}  -  \left( 1 + \frac{2k^2}{3\omega^2}\right)\frac{\dot R_b}{2k^2} \,\delta \overset{...}{ B} + 
\left[ \omega^2 \left(1- \frac{\tilde\rho _b }{2 k^2} \right)  + k^2(1+c_s^2)  - \frac{4 \tilde\rho _b}{3}  
- \frac{\ddot R_b}{k^2}\left(1+\frac{2k^2}{3\omega^2} + \frac{k^2r^2}{6}\right)
 \right] \delta \ddot B +  \nonumber \\
&&
+\left[  - \frac{\overset{...}{R_b}}{2k^2}\left( 1+ \frac{2k^2}{3\omega^2} + \frac{k^2r^2}{6} \right) + \frac{\dot R_b r^2}{12}
\left( \omega^2 \left(1-\frac{\tilde\rho_b}{2k^2} \right)  + k^2(1-c_s^2) + \frac{2\tilde\rho_b}{3} \right) - \frac{\dot R_b c_s^2}{2}\left( 1+ \frac{2k^2}{3\omega^2}\right) \right] \delta \dot B +  \nonumber \\
&&+\left[ c_s^2 k^2(k^2+\omega^2) - \frac{\tilde \rho_b \omega^2}{2}\left(1+\frac{4k^2}{3\omega^2}\right) - 2 c_s^2 \tilde \rho_b \omega^2 \left(1+\frac{2k^2}{3\omega^2}\right) 
\right] \delta B  = 0\,.
\label{deltaB-4dots-lin}
\ee

{ Keeping both $k$ and $r$ in this equation looks eclectic.  Nevertheless, it can be done in the adiabatic limit, as it is discussed above. 
Moreover, the limit of small $kr$ makes sense as well.
The applicability of this approximation 
 can be justified as follows. One sees that  Eq.~(\ref{deltaB-4dots-lin})  always
contains the combination $ (1/k^2 + r^2/6)$, where according to the arguments presented above, $r^2$ is understood
as an adiabatic variable, so we can use both $k$ and $r$. Adiabaticity is valid for $k r \gg 1$, so if we want to go to anti-adiabatic 
limit, we need to take the explicit Fourier transformation:
\be
(1/6) \int d^3 r\, r^2 e^{i {\bf k r} } \delta B(r) = - (1/6) \partial_k^2 \left[\delta B(k)\right]
\label{transform-r2}
\ee
and to compare it to $\delta B(k)/k^2$. Evidently the latter dominates at small $k$ if $\delta B(k)$ is not singular at $k\rar 0$.
If $\delta B(k)$ has a power law singularity, as $1/k^\nu$, with $\nu \lesssim 1$, then numerically the $1/k^2$ term also dominates.
We will arrive to a similar conclusion working in the coordinate space, where we need to compare $\delta B({\bf r})\,r^2/6$ with
\be 
\int \frac{d^3 k}{k^2} e^{-i{\bf k r}}\delta B({\bf k}) = \int \frac{d^3 r'}{|\bf{r - r'|}} \delta B({\bf r}) .
\label{transform-k2}
\ee
For sufficiently small $r$ the first term is again subdominant, so Eq.~(\ref{deltaB-4dots-lin}) is valid in the case of small $kr$. 
}

Next let us estimate the factor $kr$ near the Jeans wavenumber $k = k_J = \sqrt{{\tilde \rho_b}/{(2c_s^2)}}$. We have
\be
(r\,k_J)^2 = \frac{3}{2c_s^2}\frac{r_g r^2}{r_m^3} \ll 1 \quad\so\quad c_s^2 \gg r_g/r_m\,.
\label{kj-r-2}
\ee
Even for a very cold gas at $T=10$ K the speed of sound is about $c_s^2 \sim T/m_p \sim 10^{-12}$, thus the above condition~(\ref{kj-r-2}) can be fulfilled for appropriate densities. On the other hand, for extremely systems such as e.g. neutron stars the speed of sound is close to one (in units of the speed of light), and the ratio $r_g/r_m$ is also much larger. However, in such systems the background pressure is of course non vanishing, while here we study non-relativistic systems with $P_b\approx 0$, so this is already outside of the limits of validity of our assumptions.

We introduce the dimensionless time $\tau=\omega t$, and define:
\be
\label{definitions_abc}
\begin{aligned}
& \delta B \equiv z\,,\\
& R_b = - \tilde \rho_b y \,, \\ 
& a = \frac{\tilde\rho_b}{ k^2}\,,\\
& b =\frac{k^2}{\omega^2}\,,\\
& c = c_s^2\,, \\
& \chi = kr\,.
\end{aligned}
\ee

Now we can rewrite (\ref{deltaB-4dots-lin}) as the dimensionless equation:
\be
\begin{aligned}
& z'''' + \frac{a}{2}\left(1 + \frac{2b}{3}\right)y'z''' + \left[1 - \frac{a}{2}\left(1+\frac{8b}{3}\right) + b(1+c)  + a\left( 1 + \frac{2b}{3} + \frac{\chi ^2}{6}
\right) y'' \right] z'' + \\
& \qquad + \left[ \frac{a}{2} \left(1 + \frac{2b}{3} + \frac{\chi ^2}{6} \right)y''' - \frac{a \chi ^2}{12} \left(1- \frac{a}{2}\left(1- \frac{4b}{3}\right) + b(1-c)\right) y' 
+ \frac{a}{2}\left(1 + \frac{2b}{3}\right)\, bc\, y'   
\right]z'  \\
& \qquad + \left[bc(1+b) - \frac{ab}{2}\left(1 + \frac{4b}{3}\right) - 2abc\left(1+\frac{2b}{3}\right)\right]z = 0\,,
\end{aligned}
\label{z-4order} 
\ee
where a prime denotes derivative with respect to $\tau$.

We can further simplify the notation by introducing the following parameters:
\begin{subequations}
 \label{definitions_Omega_mu}
\begin{align}
\alpha &= \frac{a}{2}\left(1+\frac{2b}{3}\right)\,, 
\label{alpha}\\
\Omega^2  &= 1 - \frac{a}{2}\left(1+\frac{8b}{3}\right) + b(1+c)\,,
\label{omega} \\
\mu & = b \left[ c(1+b) - \frac{a}{2}\left(1 + \frac{4b}{3}\right) - 2ac\left(1+\frac{2b}{3}\right) \right].
\label{mu}
\end{align}
\end{subequations}

We will make no assumption on the maximum value of $k$, hence on the maximum value of $b$ which could be of order unity or even larger, but we will usually consider $k \gtrsim \tilde\rho/c_s^2$ and $c_s^2 \ll 1$, corresponding to non-relativistic sound speed and wavenumbers at least of the order of the Jeans scale, for which GR predicts no exponential growth. This results in $a\ll 1$, $c \ll 1$, so in numerical calculations we keep only first order terms in these parameters. For the moment, we will put no restrictions on $b$ and $\chi$. 

Let us rewrite~(\ref{z-4order}) in the form convenient for qualitative and quantitative analysis: 
\be
z'''' + \alpha y' z''' + \left[\Omega^2 + \left(2 \alpha +\frac{a\chi ^2}{6} \right) y'' \right] z''  + 
\left[ \left(\alpha + \frac{a\chi ^2}{12} \right) y''' -  \frac{a\chi ^2}{12} (1+b) y' \right] z' + \mu z = 0\,.
\label{eq_z-4order-abc}
\ee
Since the physically interesting quantity is the magnitude of density perturbations, we present $\delta \rho/\rho_b$ 
expressed through $z \equiv \delta B$ (e.g. in the limit of small $\chi $):
\be
\frac{ \delta \rho}{\rho_b}=z\left[\frac{1+b}{a (1+2b/3)} - 2\right] + \frac{1}{2}\,z'y' + \frac{z''}{a (1+2b/3)}\,.
\label{delta-rho}
\ee  
According to Eq.~(\ref{rho-of-rg}), when $k$ is close to its Jeans value $a=\tilde \rho_b/k^2 \sim c_s^2$ and the first term in the square brackets dominates if $c_s^2 < 1/2$, which is surely the case for non-relativistic systems.

\section{Modified Jeans Instability}
\label{sec-Jeans}

First of all, let us compare our equation~(\ref{eq_z-4order-abc}) with the usual Jeans equation (remember the definition $\tilde\rho = 8\pi G \rho$)
\be\label{Jeans_original}
\delta\ddot\rho + \left(c_s^2k^2 - \frac{\tilde\rho_b}{2}\right)\delta\rho = 0\,.
\ee
As is well known, the sign of the term $c_s^2 k^2 - \tilde\rho_b/2$
determines the stability of solutions, which are sound waves for $c_s^2 k^2 > \tilde\rho_b/2$ and unstable modes in the opposite regime. These unstable modes appear at the Jeans scale
\be
k^2 \leq (k_J^\text{GR})^2 \equiv \frac{\tilde\rho_b}{2c_s^2}\,.
\ee
In modified gravity the condition of stability is determined by the sign of $\mu$ (\ref{mu}). In the limit of small amplitude of curvature oscillations or very low frequency, we can neglect $y(\tau)$ in Eq.~(\ref{eq_z-4order-abc}) so that it is reduced to a simple equation with constant coefficients which
is solved by the substitution $z = \exp ( \gamma \tau )$. The eigenvalue $\gamma$ is thus determined by the algebraic equation:
\be
\gamma^4 + \Omega^2 \gamma^2 + \mu = 0\,,
\label{gamma-eq}
\ee
where $\Omega^2$ is given by Eq.~(\ref{omega}). The eigenvalues $\gamma^2$ solving this are:
\be
\gamma^2 = - \frac{\Omega^2}{2} \pm \sqrt{ \frac{\Omega^4}{4} - \mu}.
\label{gamma-2}
\ee
If $\mu < 0$, then one of the roots $\gamma^2 > 0$, so one of the eigenvalues is positive. It corresponds to the 
usual exponential Jeans instability, though the values of the Jeans wave vector in modified gravity and in GR are different. The magnitude of the Jeans wave number is found from the condition $\mu = 0$, which in the case of small sound speed ($c_s^2 \ll 1$) yields:
\be
a=\frac{2c (1+b)}{1+ 4b/3}\,.
\label{a-of-b}
\ee
This is an equation quadratic with respect to the Jeans wave number $k_J^\text{MG}$ [see the definitions~(\ref{definitions_abc})]. We present an explicit solution for large $\omega^2 \gg k^2,\tilde\rho$:
\be
(k_J^\text{MG})^2= (k_J^\text{GR})^2\left[1 + \frac{(k_J^\text{GR})^2}{3\omega^2} \right]\,,
\label{k-J-MG}
\ee
which recovers the GR result in the limit $\omega \rar \infty$. For large $\omega$, the physical eigenvalues corresponding to~(\ref{gamma-2}) are
\be\label{eq_Gamma_mod_grav}
\Gamma_\text{MG}^2 = \omega^2\gamma^2 \simeq
\begin{cases}
\cfrac{\tilde\rho_b}{2} - c_s^2k^2 + \cfrac{\tilde\rho_b}{2}\left(\cfrac{\tilde\rho_b}{2k^2} - c_s^2 + \cfrac{k^2\tilde\rho_b}{6\omega^2}\right)\,,\\\\
- \omega^2 - k^2 + \cfrac{\tilde\rho_b\omega^2}{2k^2}\,. 
\end{cases}
\ee
The former value gives the usual Jeans growth, up to a small correction (the term in brackets), whereas the latter corresponds to scalaron oscillations with frequency approximately equal to $\omega$, as we expected since the theory possesses an additional massive scalar mode. 

Equation~(\ref{k-J-MG}) shows that in modified gravity the Jeans wavenumber is larger than in GR. This corresponds to a reduced minimum length scale associated to structure formation. The correction is typically small, but for models in which $k_J^\text{GR}/\omega$ is non negligible it could lead to significant corrections. The phenomenology of this result and its implications for constraining models will be dealt with elsewhere.

If $\mu $ is positive, but $\mu< \Omega^4/4$, both possible values of $\gamma^2 $ are real and negative, so $\gamma$ is purely imaginary which corresponds to acoustic oscillations. Thus these two cases of negative and positive (but not too large, see below) $\mu$ are in a one-to-one correspondence to the usual Jeans analysis. The results of this paper generalize those of~\cite{ADR-Jeans}, where we analyzed gravitational instability in modified gravity over a time-independent background.

If $\mu$ is large, i.e. $\mu \geq \Omega^4/4$, there would exist a new type of unstable oscillating solution with exponentially rising 
amplitude. Indeed, $\gamma ^2$ becomes a complex number and two of the four solutions of Eq.~(\ref{gamma-eq}) for $\gamma$ 
have positive real parts. However, such solution is absent in the model studied here; using~(\ref{definitions_Omega_mu}), 
we find that the condition $\mu \geq \Omega^4/4$ becomes:
\be
\left[\frac{(4a+3c)^2}{9} + 1-2c \right]b^2 - \left[\frac{a(5-4a-21c)}{3} - 2 +2 c\right]b + \frac{(2-a)^2}{4} \leq 0\,.
\ee
This equation has a real positive solution for $b$ if and only if the coefficient $ [a(5-4a-21c)/{3} - 2 + 2c] $ is positive. 
However, this is not  fulfilled for any positive $a$ and $c$. On the other hand, it is not excluded that some other 
modified gravity models may possess such a kind of instability.

\section{Parametric Resonance}
\label{sec-param_res}

In this section we show that even for $\mu > 0$, corresponding classically to the stable (sound-wave) regime, Eq.~(\ref{eq_z-4order-abc}) may admit unstable solutions, through a mechanism analogous to parametric resonance.

The usual textbook example of parametric resonance is the Mathieu equation (for details, see e.g.~\cite{Landau-Lifshitz-1}):
\be\label{eq_Mathieu}
\ddot f(\tau) + \Omega_0^2\left[1 + h\,\cos (\Omega_1 + \epsilon)\tau\right]f(\tau) = 0\,,
\ee
where it is assumed that $h \ll 1$. If $\Omega_1/\Omega_0 = 2$, the solution grows exponentially, behaving approximately as:
\be
f(\tau) \sim \sin(\Omega_0 \tau+ \varphi)\,\exp(\gamma \tau)\,,
\ee
with\footnote{The parameter $\gamma$ introduced here is evidently different from the growth rate $\gamma$ characterising the Jeans instability, considered in the previous section.}
\be
\gamma = \frac{h\,\Omega_0}{4}\,.
\ee
Parametric resonance is excited for
\be
|\epsilon| < \frac{h\,\Omega_0}{2}\,.
\ee
There are other resonance modes, which are generated at the frequencies $\Omega_1/\Omega_0  = 2/n$, 
where $n$ is an integer, but they are usually noticeably weaker.

In modified gravity, we have derived the 4th order equation~(\ref{eq_z-4order-abc}) governing the evolution of scalar perturbations over an oscillating background, which enters through the function $y(\tau)$. An effect analogous to parametric resonance appears in this situation as well. A similarity of parametric resonance in the fourth-order equation and in the classical Mathieu equation can be seen both in the limit of high and low frequencies. In the former case one can neglect lower order derivatives in Eq.~(\ref{eq_z-4order-abc}), keeping only $z''''$, $z'''$, and $z''$, effectively reducing the equation to a second order one in the variable $z''$. In the case of low frequencies, on the other hand, $z''''$ and $z'''$ can be neglected and we come again to the Mathieu equation. In what follows we study the complete fourth order equation both analytically and numerically and observe resonant amplification.

We assume that $y(\tau)$ is a periodic function describing curvature oscillations and consider two possible forms of them:
\begin{enumerate}
\item purely harmonic ones~(see Sec.~\ref{sec-harmonic}):  
\be
y_{harm}(t)= y_{eq}(t) + y_0\cos(\omega_1 t + \theta)\,,
\label{y-harm}
\ee
where $y_{eq}$ is the equilibrium point of the potential around which the curvature oscillates, i.e. the external energy density in units of the initial density $y_{eq} = \rho(t)/\rho_0$. The harmonic solution $y_{harm}$ was found in our paper~\cite{ADR-1} when the amplitude of curvature oscillations is small. In this case $\omega_1 = \omega$ and $y_0<y_{eq}$. The time variation of $y_{eq}$ and $y_0$ is much slower than the oscillations, namely $\dot y_{eq}/y_{eq} \ll \omega_1$ and similarly for $y_0$. 

\item spiky solutions found in Ref.~\cite{ADR-1} (see Sec.~\ref{sec-spikes}), which, up to a slowly changing term, we approximate as 
\be
y_{sp}(t)=\frac{y_0\,d^2}{d^2 + \sin^2(\omega_2 t + \theta)}\,,
\label{y-sp}
\ee
where $d\ll 1$, so that we have spiky solutions (narrow peaks with a large separation between them).

According to~\cite{ADR-1} we have $\omega_2=\omega/2$, where $\omega $ is given by Eqs.~(\ref{omega-F}) and (\ref{omega-2}). 
The Fourier transform of $y_{sp}(t)$ contains modes with much higher frequencies than $\omega_2$. The dominant mode, i.e. the mode with the largest amplitude, is excited at $2\omega_2$. 

\end{enumerate}

When studying the spiky solutions, and in general any non-harmonic behaviour of the background curvature, we will have several modes with frequency at least equal to $\omega$ and possibly much larger. The overall effect can be seen as the combination of the 
individual contributions from the different modes, so it is useful to keep $\omega_1$ or $\omega_2$ as a free parameter.  
If resonance is excited for some value of $\omega_{1,2}$ and some non-harmonic solution for $R$ contains such frequency, then we can expect resonant behaviour in the full non-harmonic case too, though perhaps slightly suppressed.

\subsection{Harmonic Oscillations}
\label{sec-harmonic}

Let us start from harmonic oscillations, Eq.~(\ref{y-harm}).  
We assume that curvature oscillates with a fixed frequency. Indeed, this is a rather general scenario in $F(R)$ theories, at least when the
conditions~(\ref{approx_conditions}) and $\dot \omega/\omega^2 \ll 1$ hold.\footnote{If the frequency changes with time, the resonance effect may possibly survive
but with a changing frequency of the signal as well.}
In terms of dimensionless time, $\tau = \omega t$, $y_{harm}(\tau)$ takes the form: 
\be\label{y_harmonic}
y_\text{harm}(\tau) = y_{eq}(\tau) + y_0 \cos(\Omega_1 \tau + \theta)\,,
\ee
where $\Omega_1 = \omega_1/\omega$, $y_0$ is the amplitude of oscillations and $\theta$ is a constant phase. This form of $y$ physically corresponds to an oscillating scalar curvature~[see~(\ref{definitions_abc})]:
\be
R_b(t) = -\tilde\rho_b [y_{eq}(t) + y_0 \cos( \omega_1 t + \theta)] = R_\text{GR}[y_{eq} + y_0 \cos( \omega_1 t + \theta)] \,.
\ee
In order for this to be physically sensible we need $y_0 < y_{eq}$, so that $R$ does not change sign, but note that $y_{eq}$ 
rises with time because the background energy density is assumed to be constantly increasing, 
see e.g.  the second paper in Ref.~\cite{future-sing} and Ref.~\cite{ADR-1}. Therefore, it makes sense to explore values $y_0 > 1$.

Eq.~(\ref{eq_z-4order-abc}) is reduced  to the Mathieu equation~(\ref{eq_Mathieu}), 
if the terms containing odd derivatives in Eq.~(\ref{eq_z-4order-abc})
can be neglected and $\mu \ll 1$. We can see that this is the case when $\alpha$, $a$, $b$ and $c$ are all much smaller than unity. Although this seems rather restrictive, it is actually a natural possibility as it corresponds to $k\ll \omega$, $c_s^2\ll 1$ and $k^2\gg \tilde\rho_b$. In this case the substitution  $z'' = x$ yields:
\be
x'' + \left[ \Omega^2 + 2\alpha \cos (\Omega_1 \tau) \right] x = 0\,,
\label{x-two-prime}
\ee
If $\Omega_1/\Omega = 2$, then the parametric resonance would be excited. Comparing this to Eq.~(\ref{eq_Mathieu}), we see that the  
parameter $h$ is expressed through $\alpha$ as
\be
\label{parametric_resonance_Omega_condition}
 h = 2\alpha\Omega^{-2}\,.
\ee

\subsubsection{Fundamental harmonic}

Let us consider $\Omega_1 = 1$, as is the case for harmonic oscillations with small amplitude, in which the curvature oscillates with a frequency given precisely by~(\ref{omega-F}). Then the parametric resonance condition requires
\begin{align}
\frac{1}{4}  = \Omega^2 &\equiv 1 - \frac{a}{2} + b - \frac{4ab}{3} \nonumber\\
&= 1 - \frac{\tilde\rho_b}{2k^2} + \frac{k^2}{\omega^2} - \frac{4\tilde\rho_b}{3\omega^2}\,,
\label{Omega_condition}
\end{align}
where we express $\Omega$ in terms of physical quantities, according to Eqs. (\ref{definitions_abc}), (\ref{alpha}), and (\ref{omega}) assuming $c\ll 1$.

Due to the similarity to the standard Mathieu equation, we are tempted to study how our full equation~(\ref{eq_z-4order-abc}) behaves when~(\ref{Omega_condition}) is fulfilled. Solving for $k$ gives
\be
k^2 = \frac{1}{24}\left(16\tilde\rho_b - 9\omega^2 \pm \sqrt{81\omega^4 + 256\tilde\rho_b^2}\right) \simeq
\begin{cases}
 -\cfrac{3\omega^2}{4}\\\\
 \cfrac{2\tilde\rho_b}{3}
\end{cases}
\ee
The former solution is of course unphysical because it leads to an imaginary\footnote{In inhomogeneous systems of finite size $k$ may be imaginary, as is known for propagation in waveguides.} $k$. The latter solution
\be
\label{k_res}
k^2 \simeq k_\text{res}^2 \equiv \frac{2\tilde\rho_b}{3} 
\ee
is physically sensible but lies outside the Jeans radius, where we have unstable modes even in GR; still, provided that $k_\text{res} r_m \geq 1$ so that we are still inside the cloud considered, this would result in an enhancement of structure formation starting at the specific scale~(\ref{k_res}). The implications of this will be considered elsewhere.

\subsubsection{Higher harmonics}

As mentioned earlier, when curvature does not behave precisely as~(\ref{y_harmonic}) due to anharmonic features, we will have several modes contributing to the overall effect, and these modes will in general contain higher frequencies. Therefore, let us now keep $\Omega_1$ as a free parameter and discuss parametric resonance in this case.

As in the standard case, we assume that Eq.~(\ref{y_harmonic}) is fulfilled and insert the tentative solution
\be\label{z_try}
z(\tau) \sim (A \cos\Omega\tau + B \sin \Omega\tau)\exp(\gamma\tau)
\ee
into Eq.~(\ref{eq_z-4order-abc}). The transformed equation would contain higher harmonics, which we neglect (as is done in the standard case). In order to do so, we exploit the following exact relations:
\be
\begin{aligned}
 & 2 \cos x\,\cos 2x = \cos x + \cos 3x\,,\\
 & 2 \cos x\,\sin 2x = \sin x + \sin 3x\,,\\
 & 2 \sin x\,\cos 2x = -\sin x + \sin 3x\,,\\
 & 2 \sin x\,\sin 2x = \cos x - \cos 3x\,,
\end{aligned}
\ee
and neglect $\cos 3x$, $\sin 3x$. We also take the limit $\gamma \ll \Omega$, as expected and confirmed by results (see below). We will investigate values $\Omega_1/\Omega \simeq 2$, which as we have seen above is equivalent to the classical condition for parametric resonance.

We are left with an equation containing terms proportional to $\cos\Omega \tau$ and $\sin \Omega\tau$. We demand that both coefficient vanish simultaneously, which yields:
\be
\begin{cases}
(P_1 + P_2)A + \gamma(Q_1 + Q_2)B = 0\\
\gamma(Q_1 - Q_2)A + (P_1 - P_2)B = 0\,,
\end{cases}
\ee
where
\be
\begin{aligned}
 & P_1 = 24\mu\,, \qquad  && Q_1 = -y_0\Omega[8\Omega^2(3\alpha+a\chi^2) + (a-6\alpha)\chi^2]\,,\\
 & P_2 = y_0\Omega^2[(a\chi^2 - 6\alpha(\chi^2 + 4\Omega^2))] \qquad && Q_2 = -48\Omega^3\,.
\end{aligned}
\ee
This system admits non trivial solutions when the determinant of the associated matrix vanishes, namely when
\be
\gamma^2 = \frac{P_1^2-P_2^2}{Q_1^2-Q_2^2}\,.
\ee
The definitions~(\ref{definitions_abc}, \ref{definitions_Omega_mu}) in the limit of large $\omega$
yield the following result for the physical growth rate in the resonant case:
\be\label{eq_Gamma_res}
\Gamma^2_\text{res} = \omega^2\gamma^2 \simeq \frac{y_0^2\tilde\rho_b^2(6+\chi^2)}{576k^4}\left[(6+\chi^2)\omega^2 + (14+\chi^2)k^2\right]\simeq 
\begin{cases}
 \cfrac{y_0^2\tilde\rho_b^2\omega^2}{16k^4} & \quad  (\chi \ll 1)\,,\\\\
 \cfrac{y_0^2\tilde\rho_b^2\omega^2\chi^4}{576k^4} & \quad (\chi \gg 1)\,.
\end{cases}
\ee
The ordinary GR eigenvalue~(\ref{Jeans_original}), which at the lowest perturbation order coincides with
the modified gravity (non-resonant) solution~(\ref{eq_Gamma_mod_grav}), is
\be
\Gamma^2_\text{GR} \simeq \Gamma^2_\text{MG} \simeq \frac{\tilde\rho_b}{2} - c_s^2k^2\,.
\ee
The resonant behaviour thus dominates for
\be\label{eq_resonant_domin}
\frac{\Gamma_\text{GR}}{\Gamma_\text{res}} < 1 \qquad\so\qquad y_0^2 >  
\begin{cases}
 \cfrac{8k^4(\tilde\rho_b - 2c_s^2k^2)}{\tilde\rho_b^2\omega^2} & \qquad (\chi \ll 1)\,,\\\\
 \cfrac{288(\tilde\rho_b - 2c_s^2k^2)}{\tilde\rho_b^2\omega^2\chi^4} & \qquad (\chi \gg 1)\,.
\end{cases}
\ee
When studying the range of scales where we would have structure growth in GR ($k\leq k_J$), the factor $(\tilde\rho_b - 2c_s^2k^2)$ is positive. In the opposite region, every resonant behaviour is of course the dominant one since GR simply predicts sound waves.

Equation~(\ref{eq_resonant_domin}) indicates that there exists a lower limit for $y_0$ to activate parametric resonance, as it happens in the usual case. Nevertheless, there should be no problem in finding a sufficiently large portion of the viable parameter space in which this condition is satisfied.

\subsubsection{Numerical Results}

\paragraph{Small radii.\\}

The condition $\chi = k r \ll 1$ corresponds to the innermost portion of the cloud, at distances from the centre shorter than $k^{-1}$. Although this is likely a small region, it may still be non-negligible, especially when studying structure growth at relatively large scales (that is for relatively small $k$).

\begin{figure}[ht]
\centering
 \includegraphics[width=.40\textwidth]{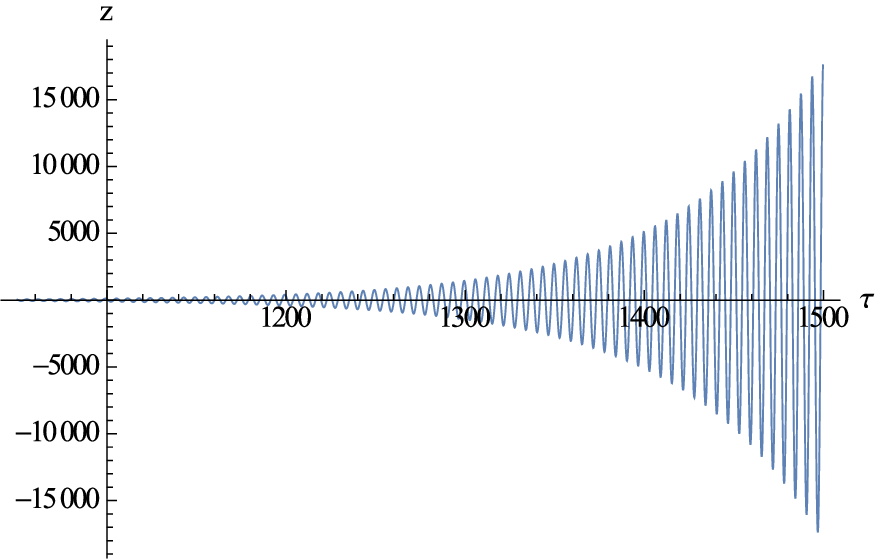}
 \includegraphics[width=.40\textwidth]{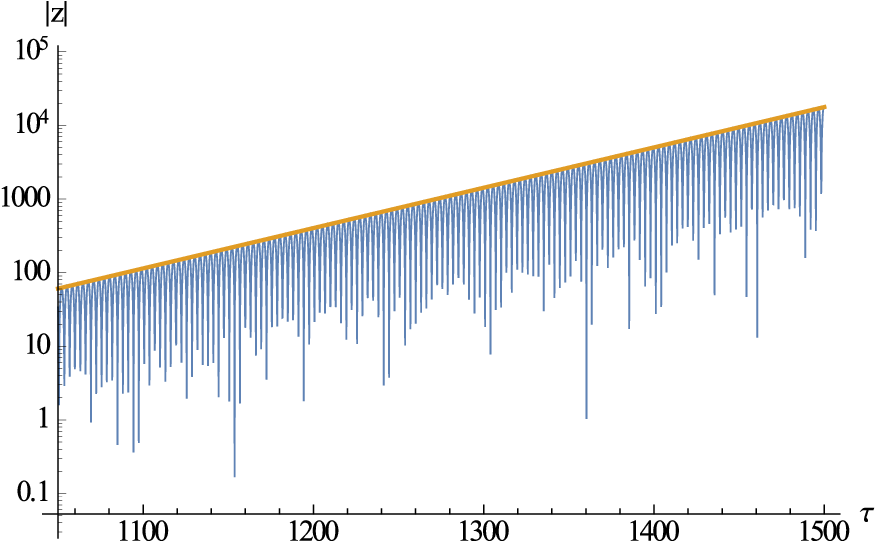}
  \includegraphics[width=.40\textwidth]{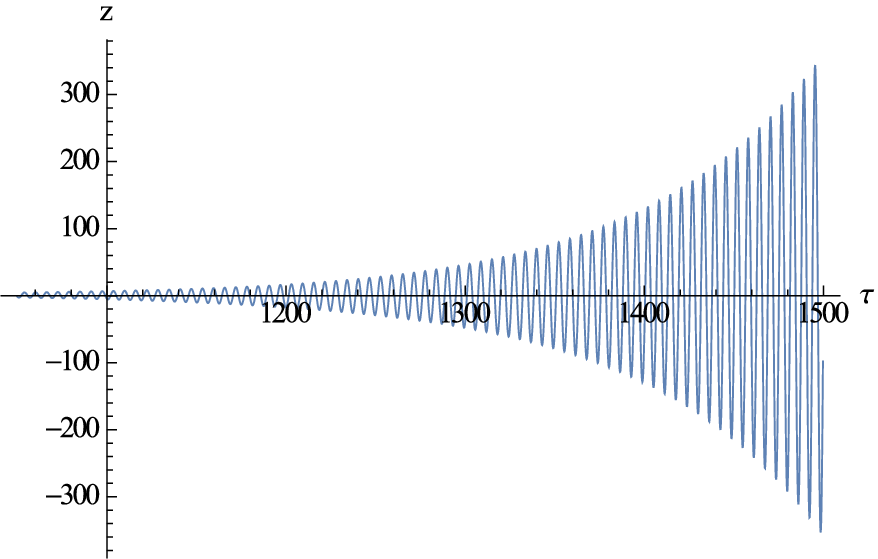}
 \includegraphics[width=.40\textwidth]{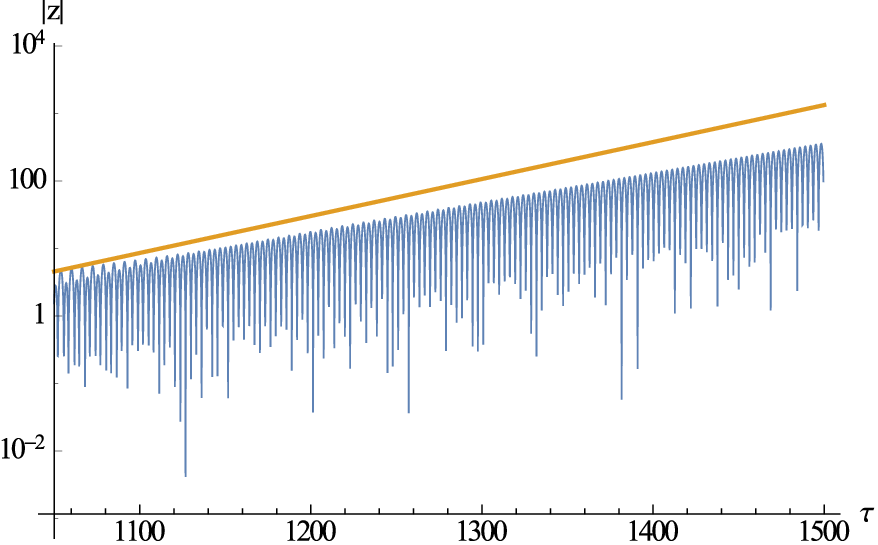}
 \caption{Solution for the parameters~(\ref{eq_params_chi_0}). 
\textit{Left panels}: Oscillations of $z$ as a function of dimensionless time $\tau$. 
\textit{Right panels}: $|z(\tau)|$ (note the logarithmic scale) and, in orange, the envelope (amplitude) of the oscillations, namely~\ref{eq_predict_exp}) with growth rate given by~(\ref{eq_Gamma_res}).\\ 
 \textit{Top}: Parametric resonance excitation of $z(\tau)$ for harmonic curvature oscillations with $y_0=5$ and $\Omega_1/\Omega=2$ and comparison with the predicted exponential rise. The relative difference between the exact and estimated growth rates is about $5\times 10^{-5}$. \textit{Bottom}: Results for $\Omega_1/\Omega = 2.015$, other parameters unvaried. Clearly the resonant behaviour is much weaker, but the agreement between the exact numerical value of $\Gamma$ and the analytical estimate (orange) is still satisfactory: $|\Gamma_\text{res} / \Gamma_\text{num} - 1 | \sim 0.24$.\\
 Note that although solutions are only shown for relatively large times, the evolution of the system starts at $\tau = 0$.\\
 }
\label{fig:harm-res}
\end{figure}

We solved Eq.~(\ref{eq_z-4order-abc}) numerically for different values of $y_0$ and $\Omega_1$. The parametric resonance 
excitation is observed at the expected frequency $\Omega_1/\Omega = 2$, see Fig.~\ref{fig:harm-res}. We present the numerical solutions for $z(\tau)$ and compare them with our analytical estimate for the amplitude of $z$, namely
\be\label{eq_predict_exp}
z \sim \exp\left(\Gamma_\text{res}t\right)\,,
\ee
with $\Gamma_\text{res}$ given by~(\ref{eq_Gamma_res}).

In this paragraphs the parameters of the medium and the wave number of the fluctuations were taken to be
\be\label{eq_params_chi_0}
a=b=\chi=0.01\,,\quad c =0.02\,,
\ee
and varying $y_0$ and $\Omega_1$. These values were merely chosen to produce the figures showing the resonant behaviour, which however appears for a very wide range of the parameters. On the other hand, parametric resonance is rather sensitive to variations of $y_0$ and of course of $\Omega_1$. In fact, the exponential growth becomes slower when we move from $\Omega_1/\Omega = 2$ towards $\Omega_1/\Omega = 2.015$ 
as one can see comparing the top and bottom panels in Fig.~\ref{fig:harm-res}.  Such sharp frequency dependence clearly 
demonstrates the resonance behaviour, with the resonance half-width roughly of the order of $\sim 0.1~\Omega$. Such a behaviour 
helps to distinguish between the parametric resonance instability and antifriction instability considered below (Sec.~\ref{s-antifric}), for which the
frequency dependence is very weak. We have not observed resonance at $\Omega_1 = \Omega$, expected in the Mathieu equation.

In both cases considered in Fig.~\ref{fig:harm-res}, the agreement of numerical calculations with the analytic estimate~(\ref{eq_Gamma_res}) 
is remarkable, and as good as $|\Gamma_\text{MG}/\Gamma_\text{num}-1|\sim 5\times 10^{-5}$ for $\Omega_1/\Omega=2$
and decreasing to about 24\% for $\Omega_1/\Omega = 2.015$, which is close to the the resonance threshold. 

\vspace*{1em}

\paragraph{Large radii.\\}

The case of large $\chi$ or $r \gg k^{-1}$ corresponds to a larger volume of the collapsing cloud. 
According to our assumptions, perturbations normally vary on scales much shorter than the total size of the object $r_m$, hence $k r_m \gg 1$. 
If we take e.g. $ 0.5\, r_m \leq r \leq r_m$, this region would occupy about 0.9 of the cloud volume, and if $ 0.1\, r_m \leq r \leq r_m$, 
then the corresponding volume is 0.999 of the total volume of the cloud.

We solved Eq.~(\ref{eq_z-4order-abc}) numerically, and present our results for the following parameters:
\be\label{eq_params_chi_10}
a = b = c = 0.01\,,\quad \chi = 10\,,\quad y_0 = 1\,,
\ee
and varying $\Omega_1$.

\begin{figure}[ht]
\centering
 \includegraphics[width=.40\textwidth]{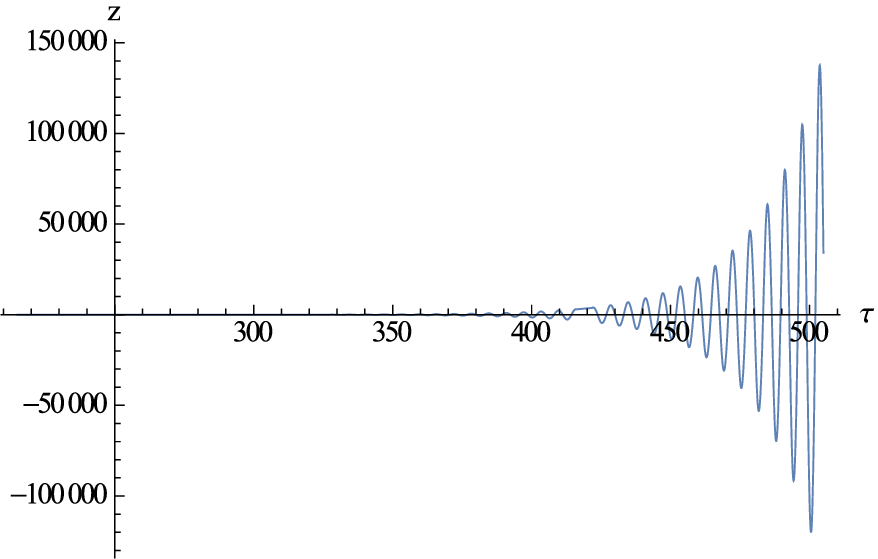}
 \includegraphics[width=.40\textwidth]{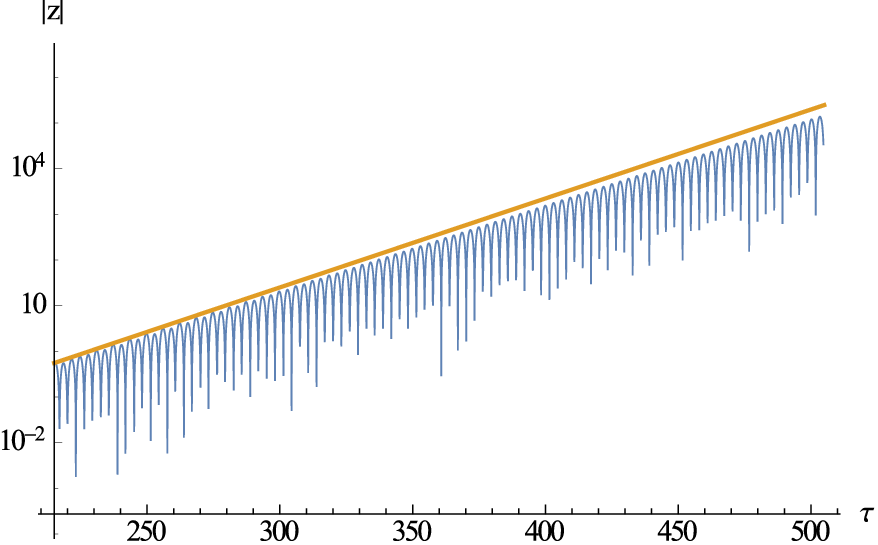}
  \includegraphics[width=.40\textwidth]{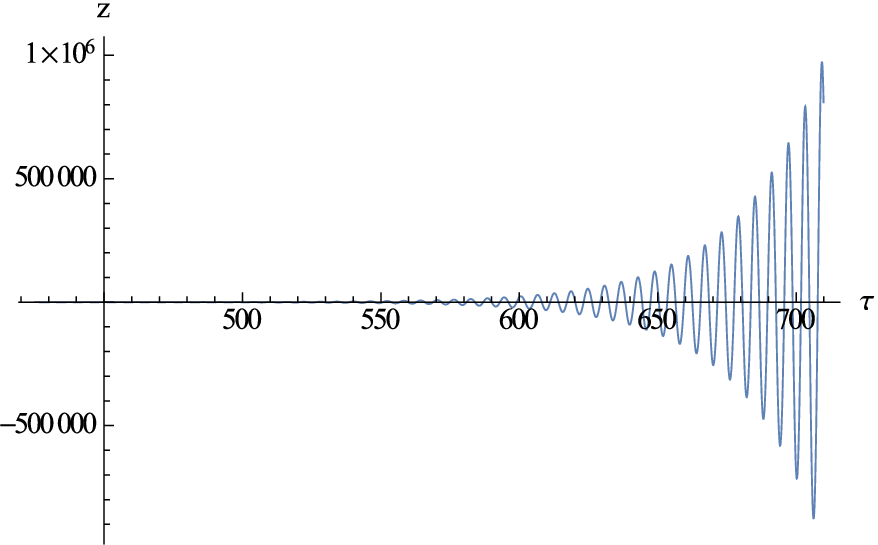}
 \includegraphics[width=.40\textwidth]{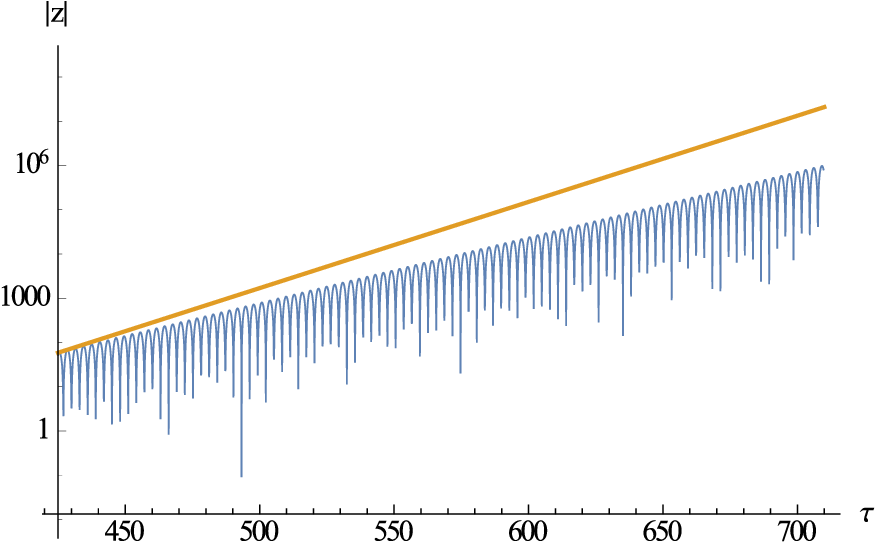}
 \caption{Solution for the parameters~(\ref{eq_params_chi_10}). Left and right panels depict the same quantities as in Fig.~\ref{fig:harm-res}.
 \textit{Top}: Solution for $y_0=1$ and $\Omega_1/\Omega=2$ and comparison with the predicted behaviour, in orange. The relative difference between the exact and estimated growth rates is $\sim 3\times 10^{-2}$. \textit{Bottom}: Results for $\Omega_1/\Omega = 2.075$, other parameters unvaried. The resonant behaviour is weaker but still rather strong, but the agreement between the exact numerical value 
 of $\Gamma$ and the analytical estimate (orange) has decreased: $|\Gamma_\text{res} / \Gamma_\text{num} - 1 | \sim 0.32$. Still, our analytical estimates are well within one order of magnitude from the numerical results.}
\label{fig:harm-res-large-chi}
\end{figure}

In Fig.~\ref{fig:harm-res-large-chi} we show results for $\Omega_1/\Omega = 2$ and $\Omega_1/\Omega = 2.075$. As we move away from the resonant value of the frequency, the agreement with our analytical estimate~(\ref{eq_Gamma_res}) decreases but remains fairly satisfactory, the relative discrepancy being much smaller than an order of magnitude. The width of the resonance region for $\Omega_1$ is roughly $\delta\Omega_1/\Omega_1 \simeq 0.05$. However, this depends strongly on the value of $y_0$: for instance, for $y_0 = 4$ the 
width is huge, about $\delta\Omega_1/\Omega_1 \simeq 1.2$. The resonance leads to a much faster growth than in the case of small $\chi$.

Of course huge values of the amplitude of oscillations, which can be easily reached numerically, do not make much sense because the first order
approximation used in deriving~(\ref{system_full}) are valid only if perturbations are much smaller than unity. With an initial 
amplitude of fluctuations of the order of $10^{-4}-10^{-5}$ (as from CMB data) the results can be trusted up to amplifications not exceeding roughly 5 orders of magnitude.

\subsection{Spike-like oscillations}
\label{sec-spikes}

As it was found in our previous works~\cite{ADR-1}, harmonic oscillations do not cover the entirety of possible solutions for the background $R(t)$. In fact, we have shown the possibility of narrow ``spikes'' of large amplitude ($|R|\gg |R_\text{GR}|$), see Eq.~(\ref{y-sp}). These spikes 
have time separation equal to $2\pi [\omega(R=R_\text{GR})]^{-1}$ and much smaller width; for instance, in the case of the model~(\ref{F-AS}), the width of the spikes can be as small as roughly $m^{-1} \leq (10^5\text{ GeV})^{-1}$. 

\begin{figure}[ht]
\centering
 \includegraphics[width=.40\textwidth]{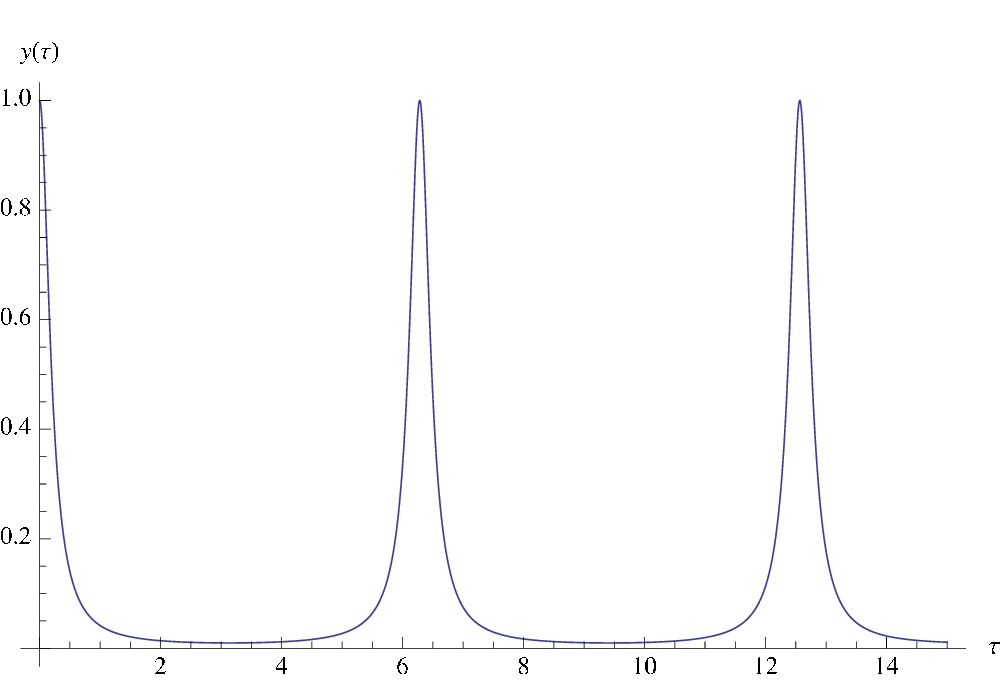}
 \includegraphics[width=.40\textwidth]{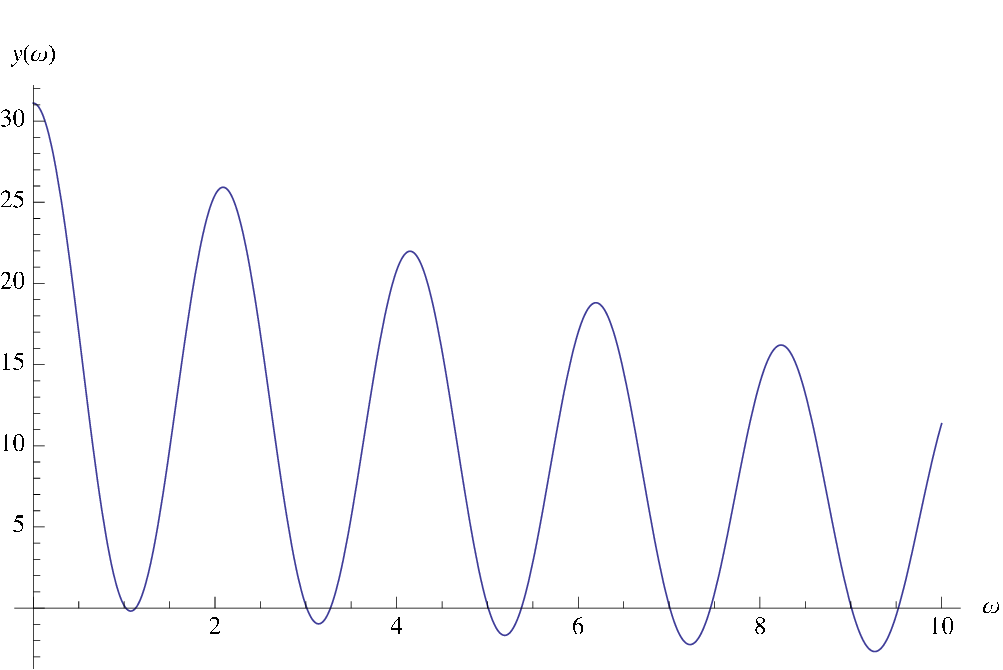}
 \caption{\textit{Left panel}: $y(\tau)$ defined by Eq.~(\ref{y-sp-tau}) with $y_0=1$, $\Omega_2=0.5$, and $d=0.1$.
\textit{Right panel}: Fourier transform of~$y(\tau)$.}
\label{fig:spikes}
\end{figure}

We solve numerically Eq.~(\ref{eq_z-4order-abc}) using $y_{sp}$ as function of the dimensionless time:
\be
y_{sp}(\tau)=\frac{y_0 d^2}{d^2 + \sin^2(\Omega_2 \tau + \theta)}\,,
\label{y-sp-tau}
\ee
with $\Omega_2 \simeq 1/2$ in agreement with the solution of Ref.~\cite{ADR-1}. This function and its Fourier transform are presented in Fig.~\ref{fig:spikes}. Note that the amplitudes of even harmonics are much larger than the amplitudes of odd ones. Moreover, the amplitudes of odd harmonics can be negative despite the fact that the curvature $y(\tau)$ remains positive, as it should.

\begin{figure}[ht]
\centering
 \includegraphics[width=.40\textwidth]{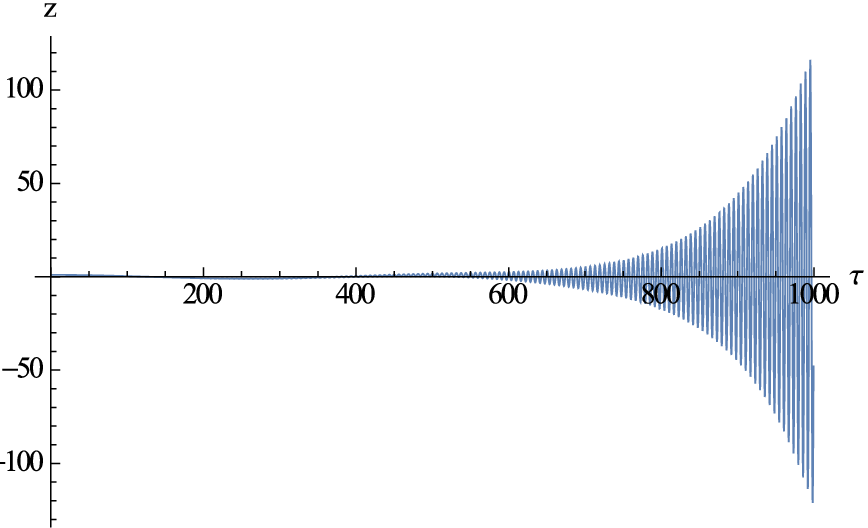}
 \includegraphics[width=.40\textwidth]{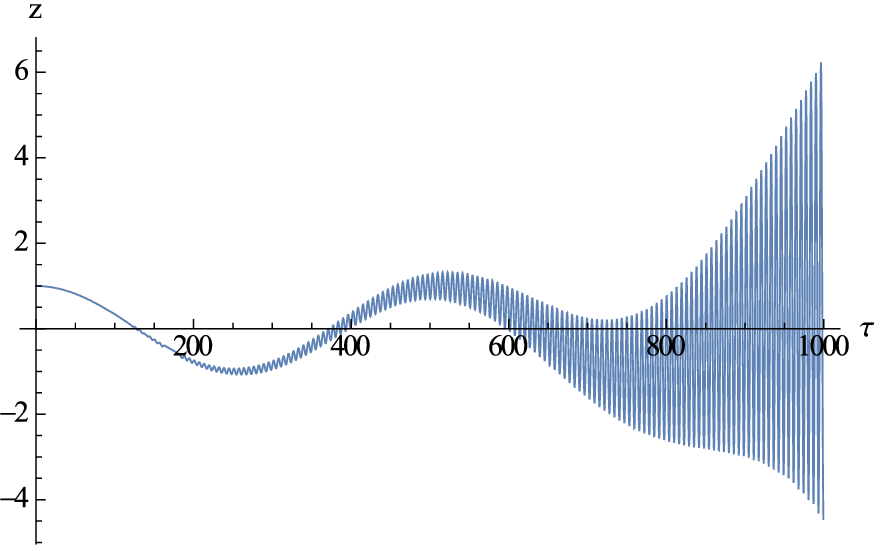}
 \caption{Parametric resonance excitation of $z(\tau)$ for spike-like curvature oscillations with $y_0=30$ and $\Omega_2/\Omega=1/2$ (\textit{left panel}) and $\Omega_2/\Omega=0.496$ (\textit{right panel}).}
\label{fig:spike-res-0-5}
\end{figure}

Fig.~\ref{fig:spike-res-0-5} (left panel) clearly shows parametric resonance at $\Omega_2=\Omega/2\simeq 0.5013$ which corresponds to the second resonant mode, see Eq.~(\ref{omega}), with the chosen values of parameters: $a=b=0.01$ and $c=0.02$. For these values $\mu$ is positive and equal to $1.47\cdot 10^{-4}$. In the right panel $z(\tau)$ is presented for a lower frequency $\Omega_2/\Omega=0.496$. We see that the resonance is still excited but much weaker. The shape of the resonance curve is symmetric with respect to the position of the resonance frequency. The sharp decrease of the
oscillation amplitude shows that the resonance is quite narrow, having width $\delta\Omega /\Omega \ll 1$.

\begin{figure}[ht]
\centering
 \includegraphics[width=.40\textwidth]{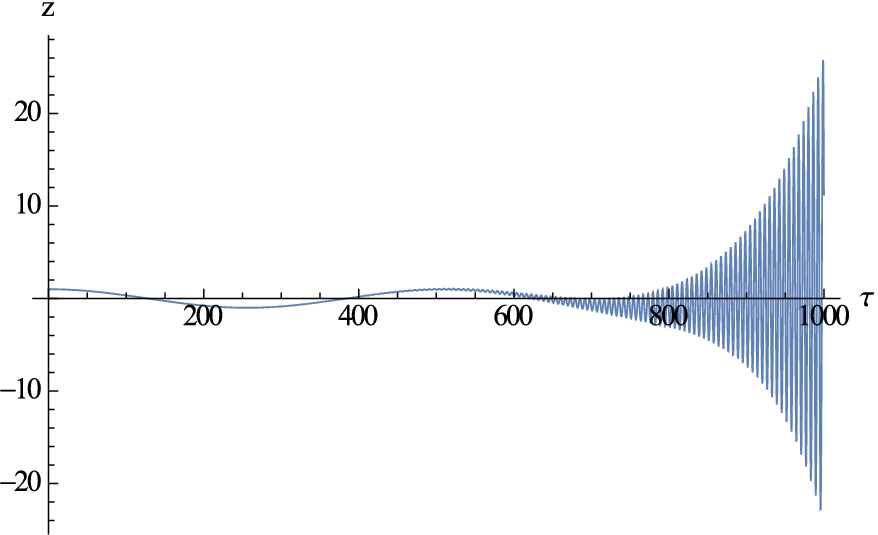}
\includegraphics[width=.40\textwidth]{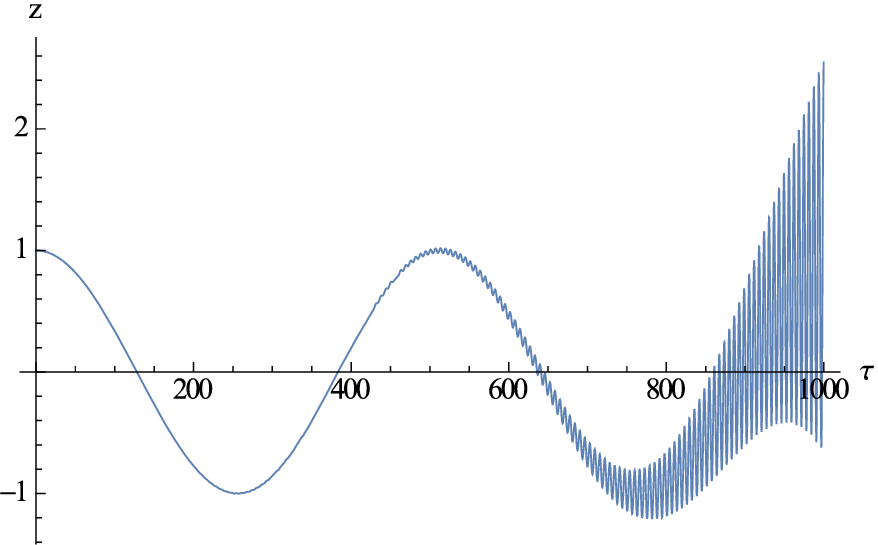}
 \caption{Parametric resonance excitation of $z(\tau)$ for spike-like curvature oscillations with $y_0=30$ and $\Omega_2/\Omega=1$ (\textit{left panel}) and $\Omega_2/\Omega=0.992$ (\textit{right panel}).}
\label{fig:spike-res-1}
\end{figure}

The main mode of parametric resonance should be at $\Omega_2/\Omega = 1$; it is presented in the left panel of Fig.~\ref{fig:spike-res-1}. However, it appears to be sub-dominant with respect to the mode at $\Omega_2/\Omega=1/2$ because of the suppression of the odd Fourier amplitudes of $y(\tau)$, see Fig.~\ref{fig:spikes}.

\begin{figure}[ht]
\centering
 \includegraphics[width=.40\textwidth]{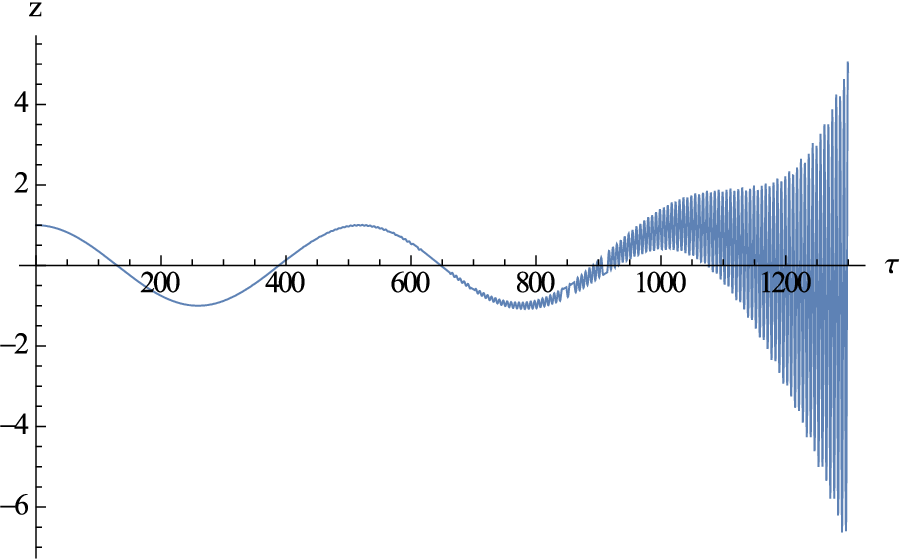}
 \includegraphics[width=.40\textwidth]{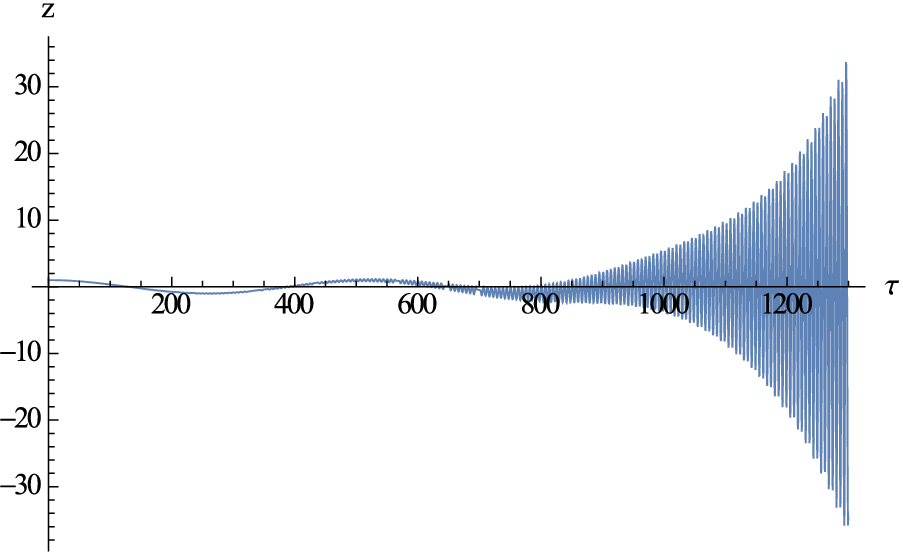}
 \caption{ Parametric resonance excitation of $z(\tau)$ for spike-like curvature oscillations with $y_0=30$ and $\Omega_2/\Omega=1/3$  (\textit{left panel}) and  $\Omega_2/\Omega=1/4$ (\textit{right panel}).}
\label{fig:spike-res-0-33-25}
\end{figure}

In Fig.~\ref{fig:spike-res-0-33-25} the evolution of $z(\tau)$ is depicted for $\Omega_2/\Omega = 1/3$ (left) and $\Omega_2/\Omega = 1/4$ (right).
These higher modes are weaker as expected, and as mentioned above the Fourier amplitudes are weaker for odd modes. We see that the 4th order equation demonstrates parametric resonance effects quite similar to the classical case~(\ref{eq_Mathieu}), though the impact of the odd derivatives in Eq.~(\ref{eq_z-4order-abc}) may be significant and lead to quantitative modifications of the results.

\section{Antifriction amplification}
\label{s-antifric}

Antifriction amplification of perturbations is induced by the change in the signs of the coefficients in front of the odd derivatives in the
4th order equation. If they become negative (and they do), then instead of damping the oscillations they would lead to their enhancement, as can be easily seen solving the oscillator equation:
\be
\ddot z + \Gamma \dot z + \omega^2 z = 0
\label{oscillator}
\ee
with $\Gamma<0$. We see below that for sufficiently large $\alpha$ (\ref{alpha}) and/or $\chi$ (\ref{definitions_abc})
the fourth order equation (\ref{z-4order}) has indeed strongly rising solutions.

\subsection{Harmonic curvature oscillations}
\label{anti-harm}

\subsubsection{Large amplitude: analytic solution}

In section~\ref{sec-Jeans} we considered the case of small amplitude of the oscillating background and saw that 
equation~(\ref{eq_z-4order-abc}) can be solved analytically. It is interesting that this equation can be also solved analytically 
in the opposite limit of large amplitude of oscillations, but with small $c$ and $\chi$.
In this case Eq.~(\ref{eq_z-4order-abc}) becomes
\be
z'''' + \alpha y' z''' + 2 \alpha z'' y'' + \alpha z' y''' = z'''' + \alpha(z'y')'' = 0\,.
\label{d4-z-large-y}
\ee
The equation is easily integrated:
\be
z''+\alpha z'y'=C_1 +C_2\tau\,,
\label{z-two-prime}
\ee
leading to the solution:
\be
z'=C_0\,e^{-\alpha y(\tau)} + C_1\, e^{-\alpha y(\tau)}\int_0^{\tau}d\tau_1\,e^{\alpha y(\tau_1)} +
C_2\,e^{-\alpha y(\tau)}\int_0^{\tau}d\tau_1 \tau_1 e^{\alpha y(\tau_1)}\,,
\label{dz-exp}
\ee
or equivalently
\be
z = z_0 + \int^\tau d\tau_1\left[C_0\,e^{-\alpha y(\tau_1)} + e^{\alpha y(\tau_1)}\int^{\tau_1}d\tau_2 \left(C_1 + C_2\tau_2 \right) e^{\alpha y(\tau_2)}\right]\,.
\ee

We take for definiteness $y(\tau)= y_0 \cos(\Omega_1 \tau)$. From the expression~(\ref{dz-exp}) it is clear that the derivative $z'$ is small when $\alpha y < 0$, while $z'$ is positive and large for $\alpha y > 0$, so the function $z$ remains constant or grows in the first and in the second case respectively.

\begin{figure}[ht]
\centering
 \includegraphics[width=.40\textwidth]{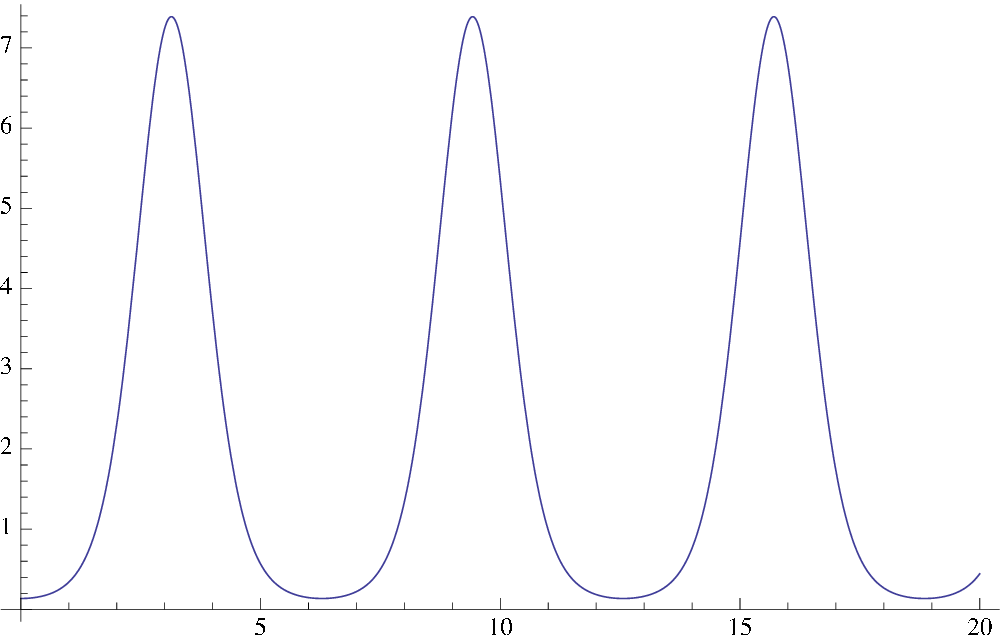}
 \includegraphics[width=.40\textwidth]{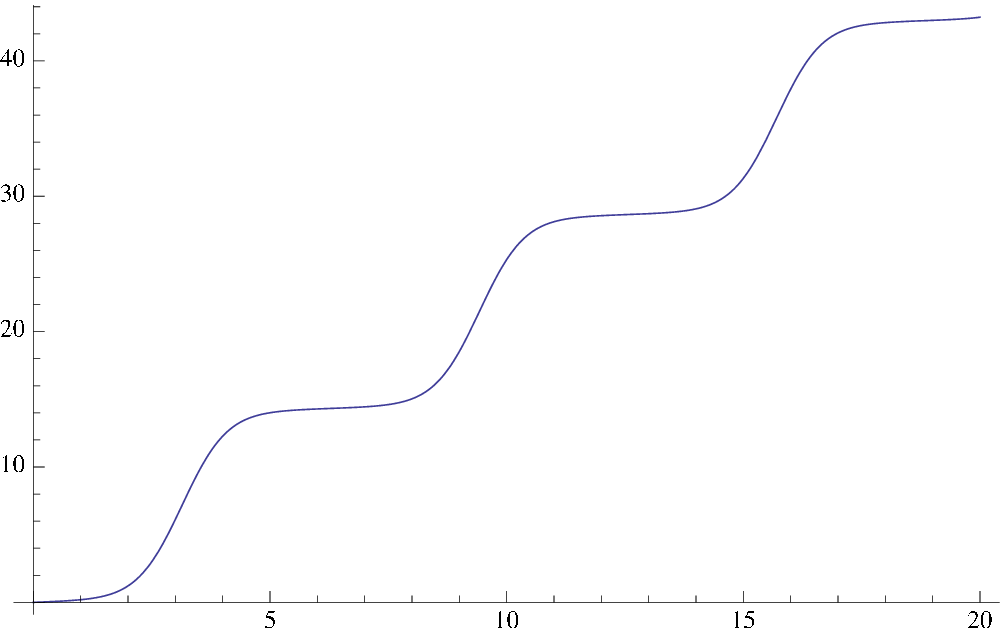}
 \caption{ The function $z(\tau)$ (\textit{right panel}) and its derivative $z'(\tau)$ (\textit{left panel}), as given by Eq.~(\ref{dz-exp}) for $C_0=1$, $C_1=C_2=0$,  $y(\tau)= y_0 \cos( \tau)$, and $\alpha y_0 =2$.  
 }
\label{fig:first-term-large-y0}
\end{figure}

\begin{figure}[ht]
\centering
 \includegraphics[width=.40\textwidth]{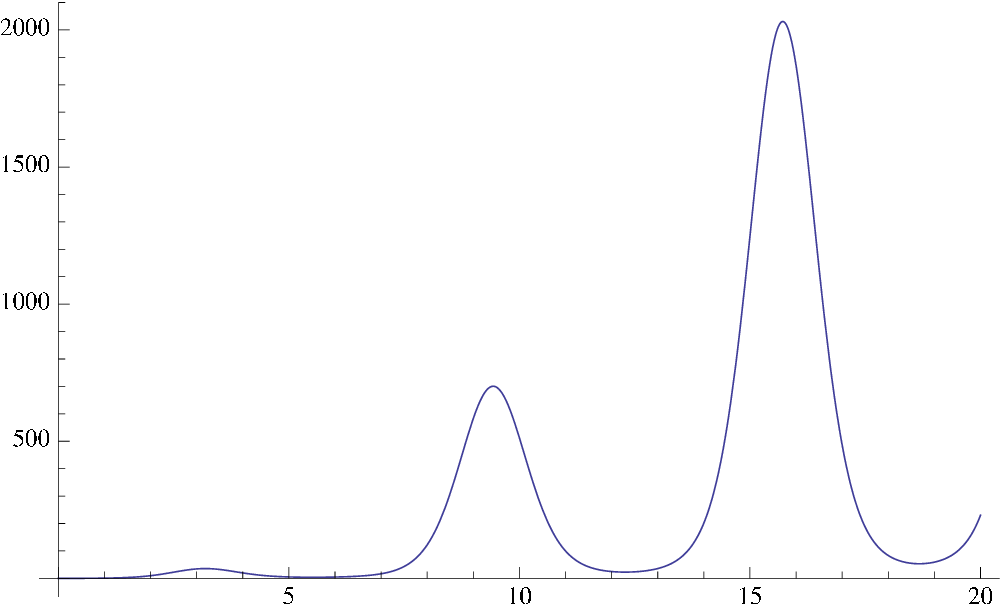}
 \includegraphics[width=.40\textwidth]{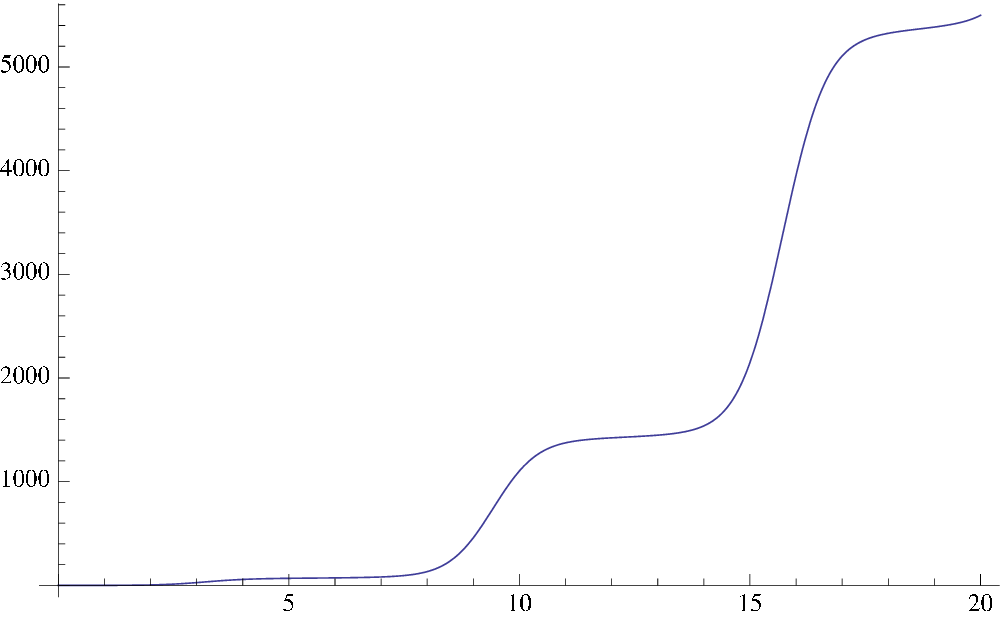}
 \caption{ The function $z(\tau)$ (\textit{right panel}) and its derivative $z'(\tau)$ (\textit{left panel}), as given by Eq.~(\ref{dz-exp}) for $C_2=1$, $C_0=C_1=0$, 
 $y(\tau)= y_0 \cos( \tau)$, and $\alpha y_0 =2$.  
 }
\label{fig:third-term-large-y0}
\end{figure}

This behaviour is shown in Figs.~\ref{fig:first-term-large-y0},~\ref{fig:third-term-large-y0}. The amplification of 
$z(\tau)$ takes place independently of the frequency of the background curvature oscillations $\omega_1$. This can be explained by the 
change of the sign of the coefficients in front of the odd derivatives in Eqs.~(\ref{eq_z-4order-abc}) and (\ref{d4-z-large-y}) 
for large $y_0$. When these coefficients are positive they act as a friction force but when they are negative they act as antifriction. Such effect is showed numerically below for the full equation~(\ref{eq_z-4order-abc}). 

\subsubsection{Numerical solutions}

\paragraph{Small radii\\}

\begin{figure}[ht]  
\centering
 \includegraphics[width=.45\textwidth]{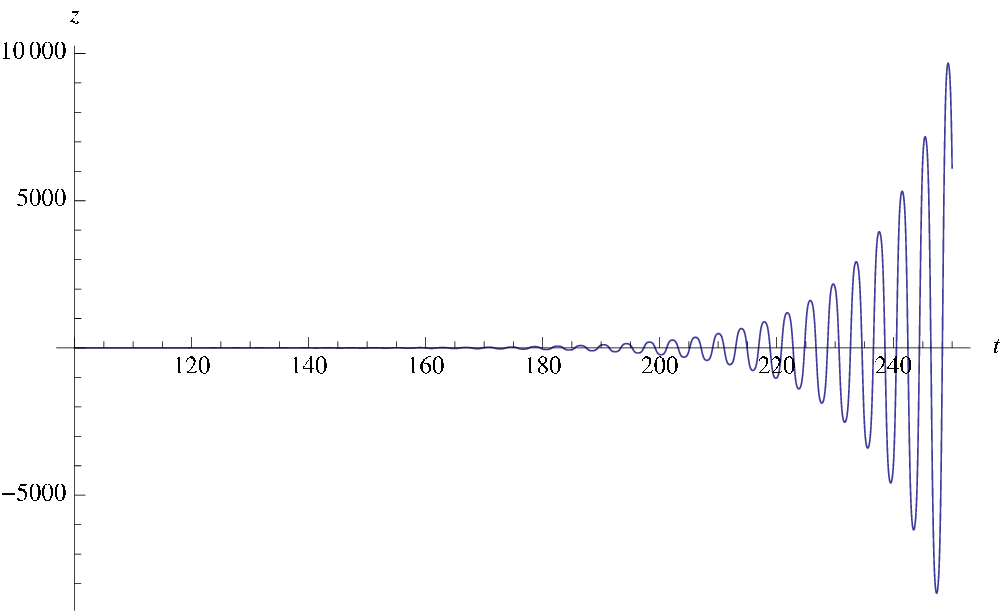}
 \includegraphics[width=.45\textwidth]{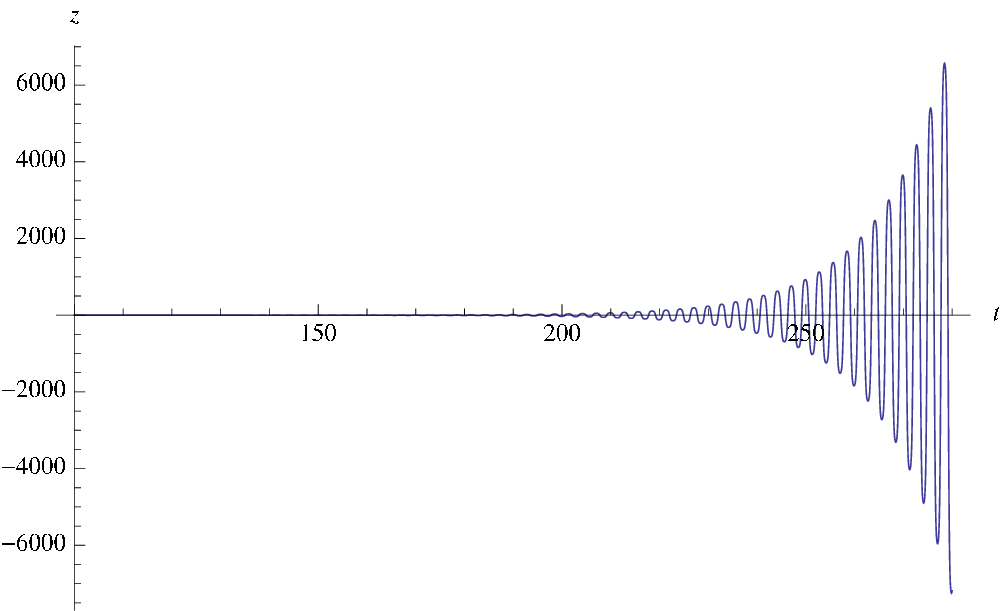}
 \caption{ Antifriction effect in the evolution of $z(\tau)$ for harmonic curvature oscillations with $y_0=169$, $\Omega_1/\Omega=3.2$  (\textit{left panel}) and  $y_0=267$, $\Omega_2/\Omega=4.4$ (\textit{right panel}). Small radii: $\chi = 0.01$.}
 \label{fig:antifric-harm-small-chi}
\end{figure}

Here we solve eq.~(\ref{eq_z-4order-abc}) numerically for the ad-hoc chosen parameters $a=0.01,\,b=0.01,\,c =0.02,\,\chi = 0.01$, for different values of $y_0$ and $\Omega_1$. The antifriction amplification is observed at the frequencies away from the resonance values, if $y_0$ exceeds a threshold value, $y_{th}$. The farther away the frequency is from the resonance, the larger the threshold. 
For example for $\omega_1 = 3.2$ the threshold value is $y_{th} = 169$, while for $\omega_1 = 4.4$ the threshold is $y_{th} = 267$. 
The evolution of $z(\tau)$ in these two cases is depicted in Fig.~\ref{fig:antifric-harm-small-chi}.

\paragraph{Large radii.\\}

A large $\chi$ leads to an antifriction instability  more efficiently, because the term proportional to $\chi^2$ enters only into the coefficients in front of the first derivative, $z'$, and there is no destructive interference of the first and third derivative terms, which in the general case may have opposite signs and act in opposite directions.

\begin{figure}[ht]  
\centering
 \includegraphics[width=.45\textwidth]{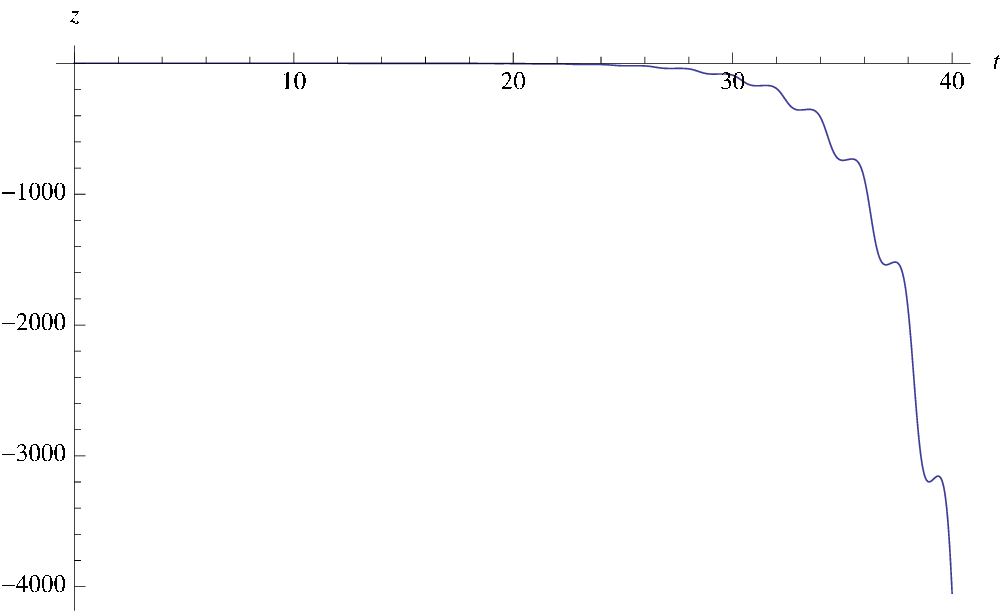}
 \includegraphics[width=.45\textwidth]{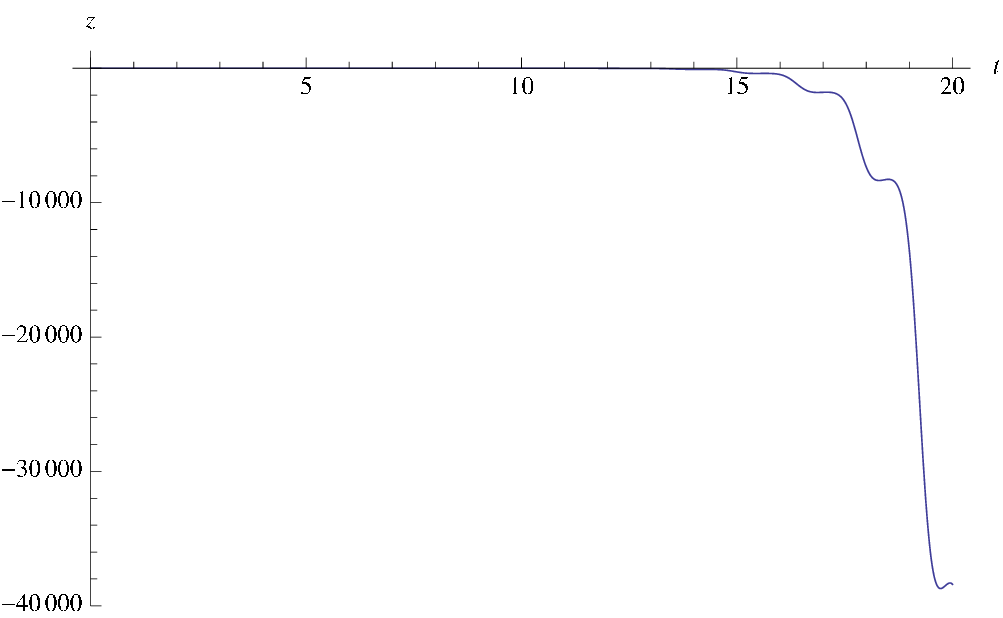}
 \caption{ Antifriction effect in the evolution of $z(\tau)$ for harmonic curvature oscillations with $y_0=10$, $\Omega_1/\Omega=3.2$  (\textit{left panel}) and  $y_0=11$, $\Omega_2/\Omega=4.4$ (\textit{right panel}). Large radii: $\chi =10$.}
 \label{fig:antifric-harm-large-chi}
\end{figure}

In Fig.~\ref{fig:antifric-harm-large-chi} we present an example of antifiriction behaviour in systems with harmonic curvature oscillations for the same values of parameters $a=0.01,\,b=0.01,\,c =0.02 $ and frequencies as in the case of small radii, except for the value of $\chi$.  We take $\chi = 10$ and again observe the threshold behaviour, but for the frequency $\Omega_1=3.2$ the threshold amplitude is $y_{th}=10$, whereas for $\Omega_1=4.4$ the threshold amplitude $y_{th}=11$.

\subsection{Spike-like curvature oscillations: numerical solutions}

\subsubsection{Small radii}

We present here the numerical solution of Eq.~(\ref{eq_z-4order-abc}) for spike-like excitations (\ref{y-sp}) with different frequencies 
$\Omega_2$ and amplitudes $y_0$. We see that for all frequencies, not only the resonant ones, the solutions are unstable, provided that $y_0$ is larger than some threshold value which depends on the frequency. We attribute this phenomenon to a change of the sign of coefficients in front of the odd derivatives in Eq. (\ref{eq_z-4order-abc}). In a sense this result is similar to that described in the previous subsection for harmonic $y(\tau)$, though quantitatively different, both in magnitude and in the shape of the signal. The latter may be explained by the different form of $y(\tau)$, by different initial conditions, or by other effects induced by a non-zero $\mu$ neglected in Eq.~(\ref{d4-z-large-y}).

\begin{figure}[ht]  
\centering
\includegraphics[width=.40\textwidth]{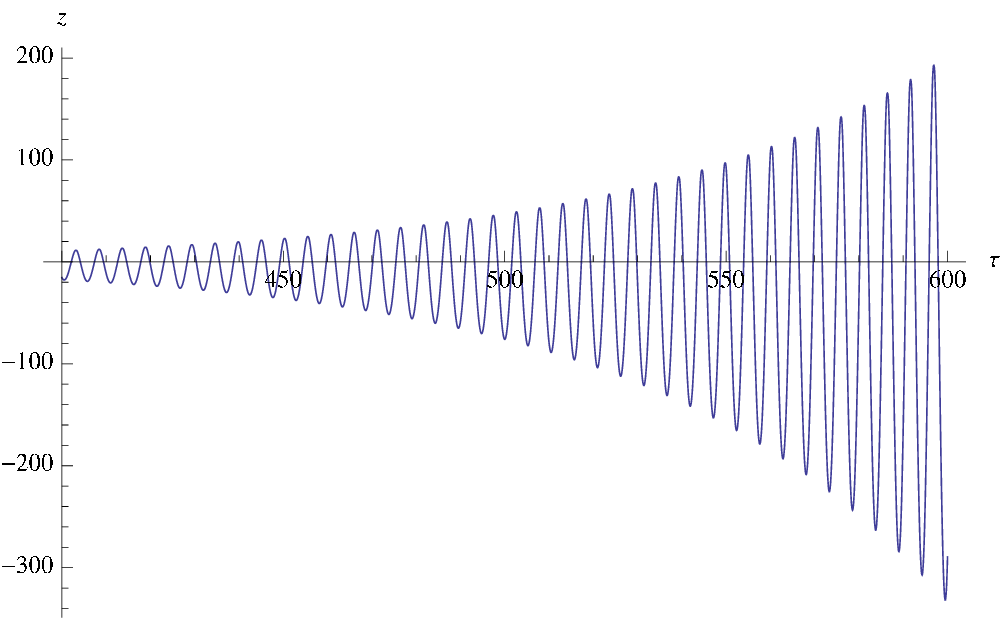}
\includegraphics[width=.40\textwidth]{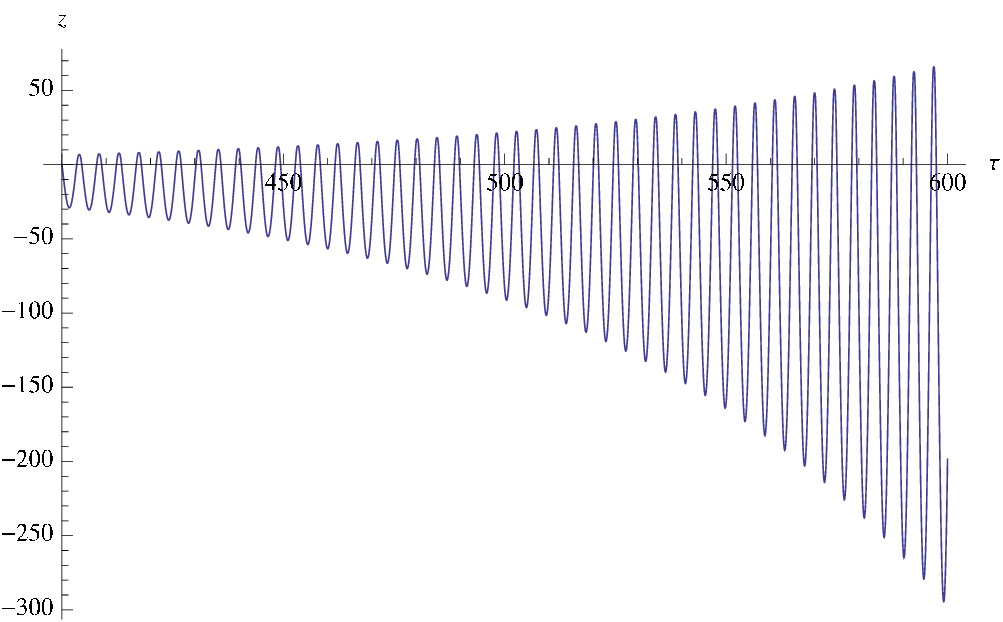}
\caption{Antifriction effect in the evolution of $z(\tau)$ for spike-like curvature oscillations with $y_0=400$, $\Omega_2/\Omega=0.6$  (\textit{left panel}) and  with  $y_0=702$, $\Omega_2/\Omega=0.7$ (\textit{right panel}).}
\label{fig:antifric-0-6-0-7}
\end{figure}

In Fig.~\ref{fig:antifric-0-6-0-7} we present the evolution of $z(\tau)$ for out-of-resonance frequencies $\Omega_2/\Omega = 0.6,\,0.7$. The farther away the frequency is from the resonant one, approximately equal to $0.5$, the larger the threshold value of $y_0$ necessary for generating an unstable solution. The magnitudes of $y_0$ taken in these figures are quite close to the threshold values, $y_0 = 400$ for $\Omega_2 /\Omega= 0.6$ and $y_0 = 702$ for $\Omega_2/\Omega = 0.7$. Note that the threshold value of $y_0$ at the resonance with $\Omega_2/\Omega \approx 0.5 $ is about 15. This threshold effect is also present in the standard parametric resonance phenomenon, if friction is non negligible \cite{Landau-Lifshitz-1}.

\begin{figure}[ht]  
\centering
 \includegraphics[width=.40\textwidth]{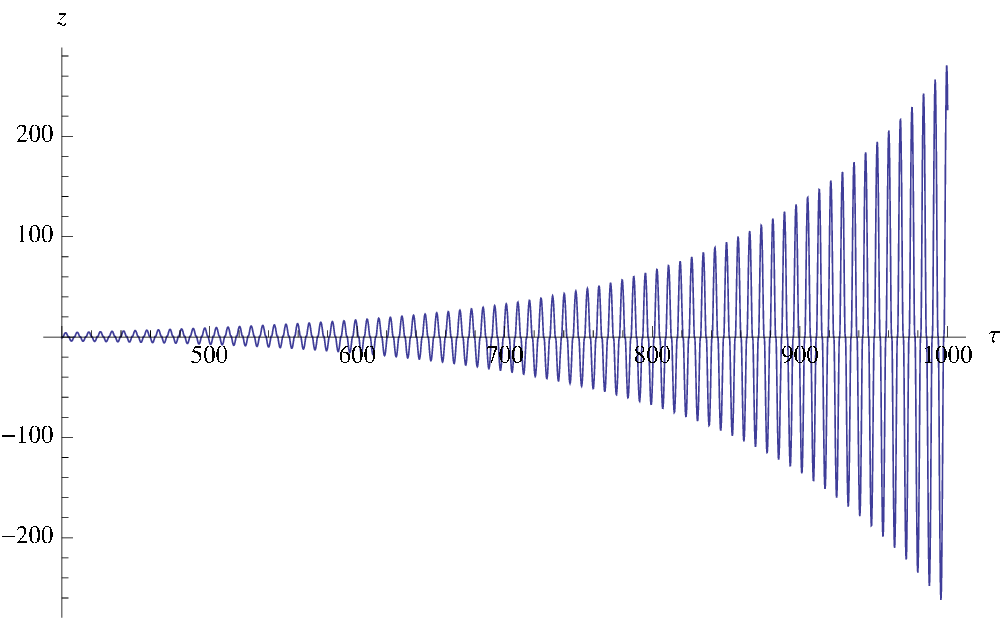}
 \includegraphics[width=.40\textwidth]{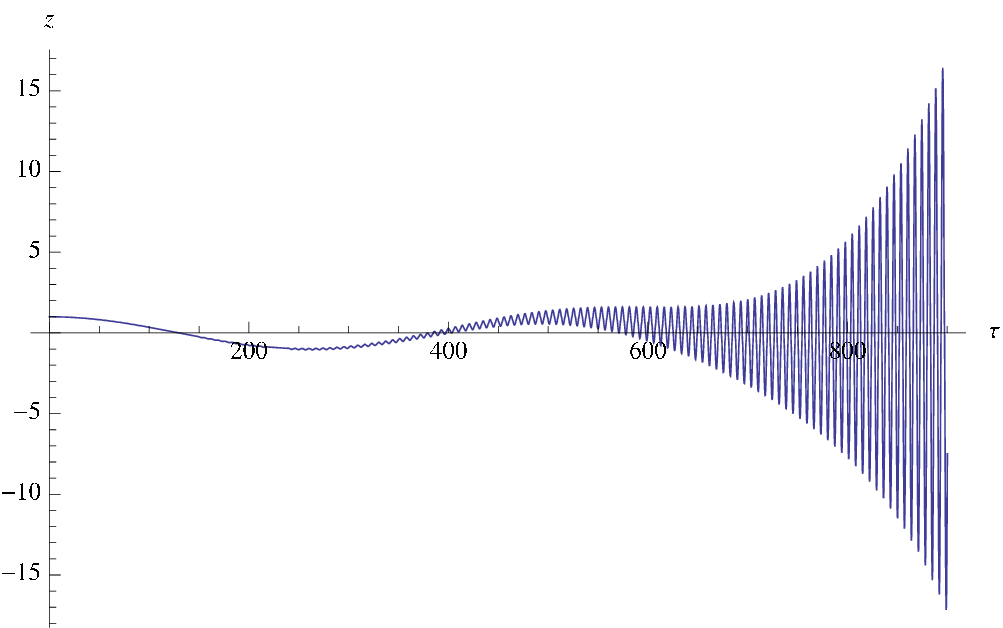}
 \caption{Antifriction effect in the evolution of $z(\tau)$ for spike-like curvature oscillations with $y_0=1280$, $\Omega_2/\Omega=0.8$  (\textit{left panel}) and  $y_0=372$, $\Omega_2/\Omega=0.9$ (\textit{right panel}).}
 \label{fig:antifric-0-8-0-9}
\end{figure}

In Fig.~\ref{fig:antifric-0-8-0-9} the evolution of $y(\tau)$ is depicted for higher frequencies $\Omega_2 /\Omega= 0.8$ (\textit{left panel}) and $\Omega_2 /\Omega= 0.9$  (\textit{right panel}) with $y_0 = 1280$ and $y_0 = 372$ respectively. The higher frequency $\Omega_2/\Omega= 0.9$ is closer to the resonant one $\Omega_2 \approx \Omega$, so the threshold value of $y_0$ is smaller. 

\subsubsection{Large radii}

\begin{figure}[ht]  
\centering
 \includegraphics[width=.45\textwidth]{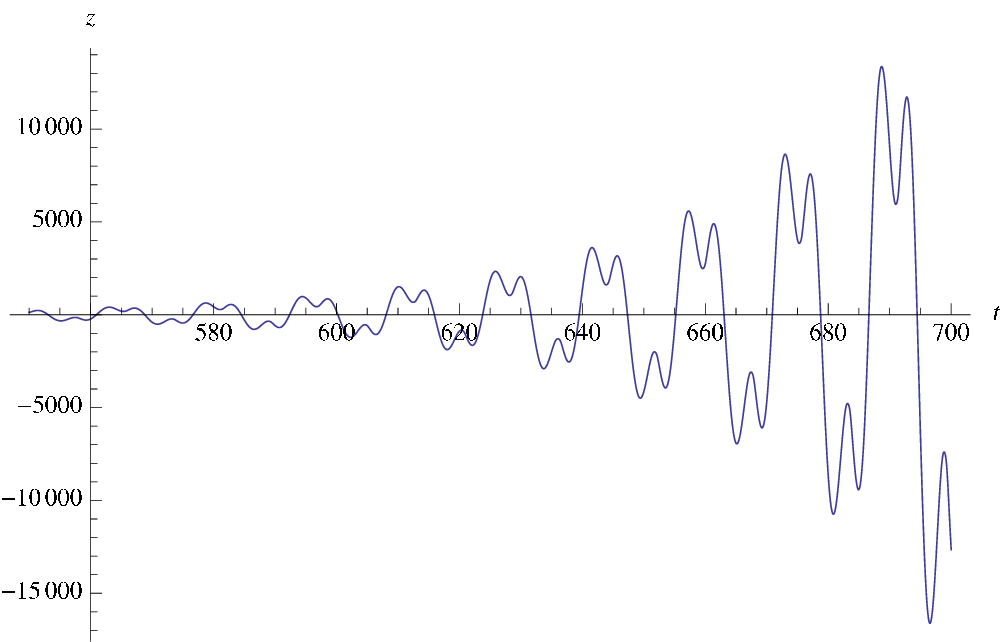}
 \includegraphics[width=.45\textwidth]{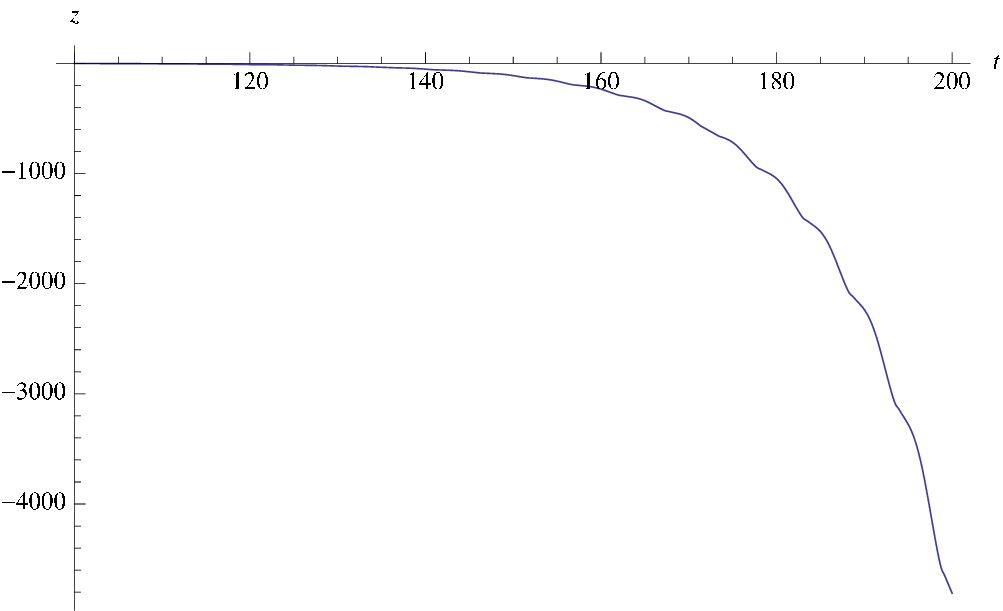}
 \caption{Antifriction effect in the evolution of $z(\tau)$ for spike-like curvature oscillations with $y_0=14$, $\Omega_1/\Omega=0.4$  (\textit{left panel}) and $y_0=9$, $\Omega_2/\Omega=0.6$ (\textit{right panel}). Large radii: $\chi =10$.}
 \label{fig:antifric-spike-large-chi}
\end{figure}

The results of numerical calculations are presented in Fig.~\ref{fig:antifric-spike-large-chi}. The instability is very clearly seen, but the character of the two kinds of instability for different values of the parameters, depicted in the left and right panels are very much different. In the first case we observe oscillations with quickly increasing amplitude, while in the second case there is an explosive, practically monotonic rise.

\section{Discussion and Conclusions} \label{s-conclusions}

A general feature of $F(R)$ modified gravity are high frequency oscillations of curvature and metric in contracting matter systems such as a gas cloud of interstellar or intergalactic matter in the process of star or galaxy formation. A similar effect might take place in the process of stellar collapse, e.g. before a supernova explosion and subsequent formation of neutron star or black hole. As we have shown in this work, the evolution of metric and density perturbations in an oscillating gravitational background possesses new and very interesting features in
addition to the usual exponential Jeans instability, which itself occurs at a different scale than in GR. The oscillating background metric and curvature of the system induce an effect analogous to the standard parametric resonance, resulting in an amplification of the fluctuations. There is a quantitative difference with respect to the standard case because the usual parametric resonance is described by a second order differential equation, while in modified gravity the equation~(\ref{deltaB-4dots-normlz}), which governs the evolution of fluctuations, is fourth order. We study here a rather general equation without restrictions to any specific $F(R)$-theory. Still, we used as a guiding example the model described by the action~(\ref{F-AS}) but did not confine ourselves to specific values of the parameters demanded by this model. We used a similar approach concerning the form of the oscillating metric and curvature background, considering the two cases of arbitrary harmonic oscillations and spike-like solutions of the type found in our previous work~\cite{ADR-1}. In both cases the evolution of perturbations is described by a fourth order equation, in which odd derivatives may play a crucial role. They describe the damping due to friction if their coefficients are positive (this is the usual case in the standard parametric resonance theory). However, such coefficients may periodically change sign and this leads to a periodic antifriction force and to a consequent amplification of perturbations. The latter effect occurs at rather high amplitude of curvature oscillations, higher than the background GR value. Systems where curvature oscillations with large amplitude are induced have been found in our work~\cite{ADR-1}.

The fourth order equations~(\ref{z-4order}) or~(\ref{eq_z-4order-abc}), which govern the evolution of the density perturbations, demonstrate a very rich pattern of different types of instabilities. There is a close analogue of parametric resonance, which is easy to describe theoretically, almost in the same way as the usual parametric resonance. However, in addition to this we found a new kind of instability induced by negative signs of the coefficients in front of the odd derivatives in the equation -- we dub it the antifriction instability. Interestingly this instability generates a completely different behaviour of the rising perturbations, depending upon the values of the system parameters. There can be an oscillating behaviour with quickly rising amplitude, or a quasi-explosive one with amplitude tending monotonically to infinity. Of course these results are reliable only if perturbations are sufficiently small, not larger than unity when written in dimensionless form.

\begin{figure}h[t]
\centering
 \includegraphics[width=.45\textwidth]{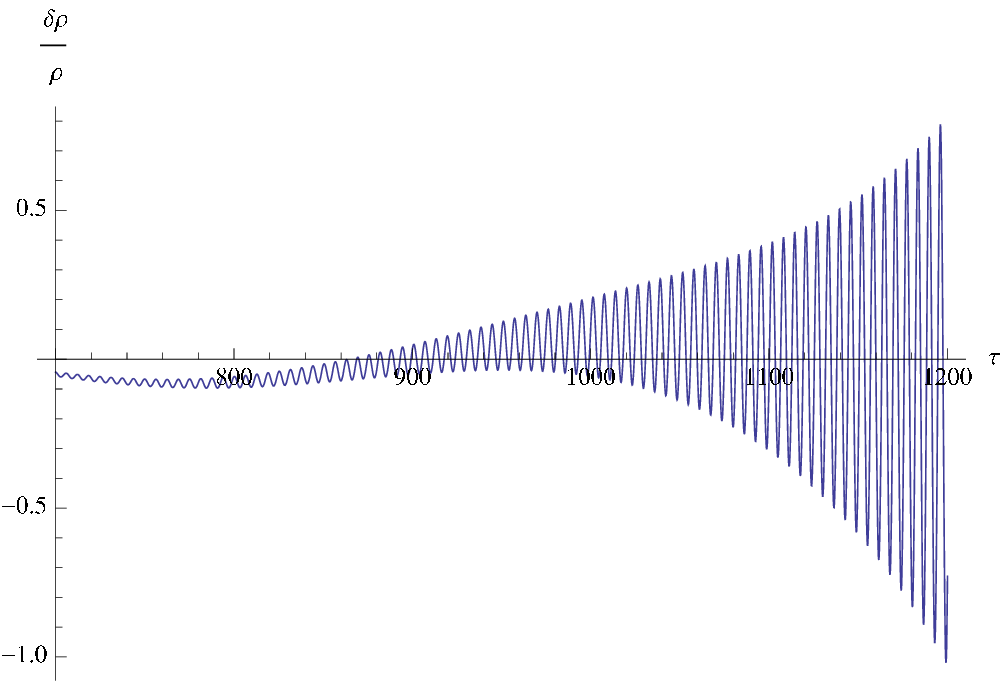}
 \caption{Evolution of $\delta \rho/\rho$ in parametric resonance region induced by the spike-like solution with $y_0=30$, $\Omega_2/\Omega = 0.5$. The initial value of metric perturbation is taken as $\delta B(0) \equiv z(0) = 10^{-3}$.}
 \label{fig:delta-rho}
\end{figure}

We have studied~(\ref{z-4order}) analytically and numerically, solving for the perturbation in the space-space metric term. Based on these solutions, one can calculate the evolution of the relative density contrast using~(\ref{delta-rho}). Barring accidental cancellations, exponentially growing solutions for $z$ will lead to an equivalent behaviour for the density perturbation $\delta\rho$, as shows in Fig.~\ref{fig:delta-rho} in the case of parametric resonance induced by the spiky solution with moderately large amplitude $y_0 = 30$.

We can see that in a relatively short time, about $10^3 \omega^{-1}$, the density perturbation would reach unity if the initial metric perturbations are  
about $10^{-3}$. The effective time can be much shorter than the gravitational time typical for the Jeans-type rise of density perturbations, since $\omega$ is usually large. Though the density contrast is an oscillating function of time, its impact on structure formation may be non-negligible and should lead to constraints on the parameters of the underlying $F(R)$-theory. This will be studied elsewhere. 

Another potentially interesting effect of the oscillating rise of curvature perturbations is the generation of curvature propagation around contracting bodies, i.e. a scalar mode for gravitational waves having mass equal to $\omega$, which may vary from practically zero up to the scalaron mass $m$. They may lead to a considerable loss of energy of the collapsing objects.

\acknowledgments
EA and AD acknowledge the support of the grant of the Russian Federation government 11.G34.31.0047.

\appendix

\section{Metric and Curvature}
\label{s-metric}

As we did previously, we consider a spherically symmetric cloud of matter with initially constant energy density inside the limit radius $r=r_m$. We choose Schwarzschild-like isotropic coordinates in which the metric takes the form:
\be
ds^2=Adt^2 - B\,\delta_{ij}\,dx^i dx^j\,, 
\label{ds-2}
\ee
where the functions $A$ and $B$ may depend upon $r$ and $t$. The corresponding Christoffel symbols are:
\be 
&&\gamma ^t_{tt}=\frac{\dot A}{2A}\,,\ \ \ \gamma ^t_{jt}=\frac{\partial_j A}{2A}\,, \ \ \ 
\gamma ^j_{tt}=\frac{ \delta^{jk} \partial_k A}{2B}\,, \ \ \ 
\gamma ^t_{jk}=\frac{ \delta_{jk} \dot B}{2A}\,, \nonumber \\
&&\gamma ^k_{jt}=\frac{ \delta^k_j \dot B}{2B}\,, \ \ 
\gamma ^k_{lj}=\frac{1}{2B}(\delta^k_l\partial_j B + \delta^k_j\partial_l B - \delta_{lj}\delta^{kn}\partial_n B)\,.  
\label{Gammas}
\ee

For the Ricci tensor, including terms quadratic in $\Gamma$'s, we obtain:
\be
R_{tt} &=& \frac{\Delta A}{2B} - \frac{3 \ddot B}{2B} + \frac{3 \dot B^2}{4B^2} + \frac{3 \dot A \dot B}{4AB} +
\frac{\partial^jA \partial_j B}{4B^2} - \frac{\partial^jA \partial_j A}{4AB}\, , \\
\label{R-tt}
R_{tj} &=& - \frac{\partial_j \dot B}{B} + \frac{\dot B \partial_jB}{B^2} + \frac{\dot B \partial_jA}{2AB}\, , \\
\label{R-tj}
R_{ij} &=& \delta_{ij} \left(\frac{\ddot B}{2A} - \frac{\Delta B}{2B} + \frac{\dot B^2}{4AB} -\frac{\dot A \dot B}{4A^2}
- \frac{\partial^kA \partial _kB}{4AB} +   \frac{\partial^kB \partial _kB}{4B^2}   \right)   \nonumber \\
&-&\frac{\partial_i\partial_j A}{2A} - \frac{\partial_i\partial_j B}{2B} +
\frac{\partial_i A\partial_j A}{4A^2} + \frac{3\partial_i B\partial_j B}{4B^2} +
\frac{\partial_i A\partial_j B + \partial_j A\partial_i B}{4AB}\, .
\label{R-ij}  
\ee  
Here and in what follows the upper space indices are raised with the Kronecker delta, $\partial^j A = \delta^{jk} \partial_k A$. 

The corresponding curvature scalar is:
\be 
R= \frac{\Delta A}{AB} - \frac{3 \ddot B}{AB} + \frac{2\Delta B}{B^2}+
\frac{3\dot A\dot B}{2A^2B} - \frac{\partial^jA\partial_jA}{2A^2B} - \frac{3\partial^jB\partial_jB}{2B^3} +
\frac{\partial^jA\partial_jB}{2AB^2}\,.
\label{R}
\ee   
Let us now present the expressions for the Einstein tensor $G_{\mu\nu}=R_{\mu \nu} - 1/2 \,g_{\mu \nu} R$:
\be 
G_{tt}&=&-\frac{A\Delta B}{B^2}+\frac{3\dot B^2}{4B^2}+\frac{3A \partial^jB\partial_jB}{4B^3}\,,  
\label{G-tt}\\
G_{tj}&=&R_{tj}\,,  \\
G_{ij}&=&\delta_{ij} \left(\frac{\Delta A}{2A}+\frac{\Delta B}{2B} - \frac{\ddot B}{A} + \frac{\dot B^2}{4AB} +
\frac{\dot A \dot B}{2A^2}
- \frac{\partial^kA \partial _kA}{4A^2} -  \frac{\partial^kB \partial _kB}{2B^2}   \right)   \nonumber \\
&-&\frac{\partial_i\partial_j A}{2A} - \frac{\partial_i\partial_j B}{2B} +
\frac{\partial_i A\partial_j A}{4A^2} + \frac{3\partial_i B\partial_j B}{4B^2} +
\frac{\partial_i A\partial_j B + \partial_j A\partial_i B}{4AB}\, .
\label{G-ij}
\ee


The energy-momentum tensor is taken in the perfect fluid form without dissipative corrections:
\be 
T_{\mu\nu}=(\rho + P)U_{\mu}U_{\nu} - P g_{\mu \nu}\,,
\label{T-mu-nu}
\ee
where $\rho $ and $P$ are respectively the energy density and pressure of the fluid and the four-velocity is:
\be
U^{\mu}=\frac{dx^{\mu}}{ds} \ \ \ {\rm and} \ \ \ U_{\mu}=g_{\mu \alpha}U^{\alpha}\,.
\label{U-mu}
\ee
We assume that the three-velocity $v^j = dx^j/dt$ is small and thus neglect terms quadratic in $v$. Correspondingly, 
\be
U_j = - \frac{B v_j}{\sqrt{A} \sqrt{1-(B/A) v_j v^j}} \approx  - \frac{B v_j}{\sqrt{A}}\,.
\label{U-j}
\ee
According to our definition $v_j=v^j$. From the condition 
\be
1=g^{\mu\nu} U_{\mu}U_{\nu} = \frac{1}{A}U_t^2 - \frac{1}{B}\delta^{kj} U_k U_j \approx  \frac{1}{A}U_t^2
\ee
we find $U_t \approx \sqrt{A}$. Now we can write:
\be
T_{tt} &=& (\rho +P) U_t^2 - P A \approx \rho A\,, 
\label{T-tt}\\
T_{jt}  &=& (\rho +P) U_t U_j \approx - (\rho +P) B v_j\,, 
\label{T-jt}\\
T_{ij} &=&  (\rho +P) U_i U_j - P g_{ij} \approx P B \delta_{ij}\,.
\label{T-ij}
\ee 

Background metric for spherically symmetric distribution of matter (an analog of the Schwarzschild solution for modified gravity) has been found in several works. We use here the form for the internal solution obtained in our paper \cite{ADR-anti} (references to other papers can be found there):
\be
B_b (r, t) & =& 1 + \frac{2M(r,t)}{\mpl^2r} \equiv  1+ B_1^{(Sch)}\,,
\label{B-of-r-t}\\ 
A_b (r,t)  &=& 1 + \frac{R_b(t )\,r^2}{6} + A_1^{(Sch)} (r,t) \label{A-of-r-t}\, ,
\ee
where 
\be
M (r,t) &=&\int_0^{r} d^3r\,  T_{00} (r,t)=4\pi \int_0^{r} dr\,  r^2\,  T_{00} (r,t)\,, \\
\label{M}
A_1^{(Sch)} (r,t)& = &\frac{r_g r^2}{2r_m^3} -\frac{3r_g}{2r_m}+
 \frac{ \pi \ddot\rho_m}{3 m_{Pl}^2}\, ( r_m^2 - r^2)^2\, ,
\label{A-1-2}
\ee
and $r_g=2M/m_{Pl}^2$ with $M$ being the total mass of the object under scrutiny.

\end{document}